\documentclass[a4paper,fleqn,usenatbib]{mnras}

\usepackage[T1]{fontenc}
\usepackage{ae,aecompl}

\usepackage{graphicx}	
\usepackage{amsmath}	
\usepackage{amssymb}	
\usepackage{threeparttable} 
\usepackage{ulem}
\usepackage{color,soul}
\usepackage[dvipsnames]{xcolor}
\usepackage{newtxtext,newtxmath}

\title[Massive core/star formation triggered by CCC]{Massive core/star formation triggered by cloud-cloud collision:\\
II High-speed collisions of magnetized clouds}

\author[Nirmit Sakre et al.]{Nirmit Sakre$^{1,5}$\thanks{E-mail: nirmitsakre@tap.scphys.kyoto-u.ac.jp (NS)},
Asao Habe$^{1}$,
Alex R. Pettitt$^{2}$,
Takashi Okamoto$^{1}$,
Rei Enokiya$^{3}$,
\newauthor Yasuo Fukui$^{4}$, and Takashi Hosokawa$^{5}$
\\
$^{1}$Department of Physics, Faculty of Science, Hokkaido University, Kita 10 Nishi 8, Kita-ku, Sapporo, Hokkaido 060-0810, Japan\\
$^{2}$Department of Physics and Astronomy, California State University, Sacramento, California 95826, United States of America\\
$^{3}$Department of Physics, Faculty of Science and Technology, Keio University, 3-14-1 Hiyoshi, Kohoku-ku, Yokohama, Kanagawa 223-8522, Japan\\
$^{4}$Department of Physics, Faculty of Science, Nagoya University, Furo-cho Chikusa-ku, Nagoya, Aichi 464-8602, Japan\\
$^{5}$Department of Physics, Graduate School of Science, Kyoto University, Sakyo-ku, Kyoto 606-8502, Japan}

\date{Accepted XXX. Received YYY; in original form ZZZ}

\pubyear{2020}

\begin{document}
\label{firstpage}
\pagerange{\pageref{firstpage}--\pageref{lastpage}}
\maketitle

\begin{abstract}
We study the effects of the magnetic fields on the formation of massive, self-gravitationally bound cores (MBCs) in high-speed cloud-cloud collisions (CCCs). Extending our previous work \citep{2021PASJ...73S.385S}, we perform magnetohydrodynamic simulations following the high-speed (20 - 40 km s$^{-1}$) collisions between two magnetized (4 \textmu G initially), turbulent clouds of different sizes in the range of 7 - 20 pc. We show that a magnetic field effect hinders the core growth, particularly after a short-duration collision during which cores cannot get highly bound. In such a case, a shocked region created by the collision rapidly expands to the ambient medium owing to the enhanced magnetic pressure, resulting in the destruction of the highly unbound cores and suppression of gas accretion to massive cores. This negative effect on the MBC formation is a phenomenon not seen in the past hydrodynamic simulations of similar CCC models. Together with our previous work, we conclude that the magnetic fields provide the two competing effects on the MBC formation in CCC; while they promote the mass accumulation into cores during a collision, they operate to destroy cores or hinder the core growth after the collision. The duration of collision determines which effect prevails, providing the maximum collision speed for the MBC formation with given colliding clouds. Our results agree with the observed trend among CCC samples in the corresponding column density range; clouds with higher relative velocity require higher column density for the formation of massive stars \citep{2021PASJ...73S..75E}.
\end{abstract}


\section{Introduction} \label{introduction}
Massive stars are important in astrophysics, since they strongly impact the interstellar medium (ISM) and galactic evolution through ejection of a large amount of energy via stellar winds and supernovae, as well as through ionization of the surrounding gas. They also supply metals into the ISM, which are then used to form the next generation of stars. However, despite their importance, the details of the formation of massive stars are still poorly understood \citep{ 2007ARA&A..45..481Z,2014prpl.conf..149T}.

For massive star formation, monolithic collapse and competitive accretion scenarios are proposed \citep{2007ARA&A..45..481Z}.
Both scenarios require a massive molecular core or massive gas region for massive star formation. It has been pointed out that a large gas accretion rate is needed to overcome the radiation pressure from a protostar embedded in its dense core in the monolithic collapse scenario of massive star formation \citep{2002Natur.416...59M}. Other processes like the disk wind can limit the mass growth of a protostar \citep{2017ApJ...835...32T}. \citet{2002Natur.416...59M} proposed that in a massive core with sufficient internal turbulent motion, a high accretion-rate flow can persist, and its ram pressure can exceed the radiation pressure from a protostar. Magnetic fields act as a support mechanism against gravitational collapse and can increase the Jeans mass, thus increasing the mass of the dense core by preventing its fragmentation \citep{Offner_2014}. 
A gas flow onto a protostar with large accretion rate can be realized in such a dense core. In this way, sufficient turbulence and magnetic fields can play an important role in massive star formation in such massive cores.
However, the formation of such massive cores is still unclear. External triggering mechanisms are frequently proposed as a promising mechanism to induce such massive core formation \citep{1998ASPC..148..150E}.

Cloud-cloud collisions (CCCs) are one of the most important candidates for triggering massive star formation \citep[see recent review by][]{2021PASJ...73S...1F}. 
The collision of molecular clouds at a supersonic speed creates a shocked region at the interface of those colliding clouds. Radiative cooling further increases the density in the shocked region. 
Dense clumps can be formed in the dense shocked region due to the enhanced self-gravity \citep{1984ApJ...279..335G, 1994MNRAS.268..291W}. Collision of clouds with different sizes would frequently occur, since the observed cloud mass function shows small mass molecular clouds are more numerous than massive molecular clouds \citep{1987ApJ...319..730S,2015ARA&A..53..583H}. A collision of clouds with different sizes creates a converging flow in the shocked region due to difference in their sizes, and this flow can grow the mass of the dense clumps as shown by \citet{1992PASJ...44..203H}. Massive cores can form in these clumps. We can expect massive star formation in these massive cores.
 
Massive star formation by CCCs is supported by numerous observational evidence \citep[e.g.,][]{1994ApJ...429L..77H,2009ApJ...696L.115F,2014ApJ...780...36F,2016ApJ...819...66D,2017ApJ...835..142T,2018PASJ...70S..46F,2018ApJ...859..166F,2021PASJ...73S.273F}. The typical observed collision speeds are between 10 to 20 km s$^{-1}$. This is much higher than the sound speed, $\sim$ 0.2 - 0.3 km s$^{-1}$, of molecular gas and internal turbulence
speeds, $\sim$ 2 - 5 km s$^{-1}$, in typical molecular clouds  \citep{1981MNRAS.194..809L,2015ARA&A..53..583H}. 

Recently, many hydrodynamic simulations of CCCs have been carried out \citep[e.g.,][]{2014ApJ...792...63T, 2015MNRAS.453.2471B,2018PASJ...70S..58T,2018PASJ...70S..54S, 2020MNRAS.496L...1D,2020MNRAS.499.1099L,2020MNRAS.494..246T}. \citet{2014ApJ...792...63T,2018PASJ...70S..58T} and \citet{2018PASJ...70S..54S} simulated CCCs to study the dense core formation, assuming some initial turbulence in the clouds. They demonstrated that dense cores formed in the shocked region of colliding clouds can grow their mass by gas accretion, if the dense cores are supported by their internal turbulence and then have time for the gas accretion.

Magnetohydrodynamic (MHD) simulations of CCCs have been carried out by several authors \citep[e.g.,][hereafter Paper I]{2013ApJ...774L..31I,2014ApJ...785...69C,2015ApJ...810..126C,2017ApJ...841...88W,2017ApJ...835..137W,2018PASJ...70S..53I,2020ApJ...891..168W,2021MNRAS.502.2285D,2021PASJ...73S.385S}. \citet{2014ApJ...785...69C} has shown that the typical mass of pre-stellar cores formed in the colliding clumps in a large molecular cloud are unaffected by the magnetic fields. 
They found the formation of low-mass cores of masses $\sim$ 0.04 - 2.5 $M_{\odot}$ in their simulations. They assumed collision speed of $\sim$ 4 km s$^{-1}$, which is as high as the typical turbulent velocity observed in molecular clouds. 
Since the typical turbulent velocity is much smaller than the observed collision speeds of CCCs, CCC simulations with higher collision speeds than their simulations should be carried out to study the collision of magnetized clouds with the observed collision speeds. \citet{2017ApJ...835..137W} and \citet{2017ApJ...841...88W} have shown that the star formation in the CCC in their numerical simulations with the typical collision speed is unaffected by magnetic fields. 
\citet{2020ApJ...891..168W} studied the effect of various magnetic field strengths on star formation in colliding clouds. 
In their strongest critically magnetized, colliding clouds model (50 \textmu G), which is near the magnetically critical case, they found a reduced number of newly formed stars due to high magnetic pressure. They used a spatial resolution of 0.1 pc which is comparable to sizes of dense cores of which the typical scale is 0.1 pc \citep{2007ARA&A..45..339B}. Internal properties of dense cores (e.g., their internal turbulent energies, magnetic field energies, and self-gravitational energies) are important for evolution of dense cores. A higher spatial resolution is required to well-resolve the dense cores than their simulations for such a study. \citet{2013ApJ...774L..31I} carried out colliding flow simulations and found that massive cores are formed due to MHD shock. \citet{2018PASJ...70S..53I} developed this study with sink particles. They simulated collision of a dense clump with uniform, dense region with a typical collision speed with a highly spatial resolution of 0.0015 pc which is much smaller than the typical size of the molecular core. Their results are in favor of massive star formation due to the effect of magnetic fields.

In Paper I, we studied the effect of magnetic fields on massive core formation in CCCs. 
We assumed turbulent clouds in uniform magnetic fields of various strengths and directions with respect to the collision direction, with assumed collision speed of 10 km s$^{-1}$. We used a minimum cell size of 0.015 pc, which is enough to resolve dense cores. 
We found a greater number of massive cores in models with a strong magnetic field of 4 \textmu G, which is due to suppression of nonlinear thin shell instability (NTSI) in the shocked region by the magnetic field, than in the very weak magnetic field (0.1 \textmu G) models. 
The strong magnetic field also supports the dense core against early gravitational collapse, suppressing the formation of low-mass bound cores, and enabling dense cores to accumulate mass more than 10 $M_{\odot}$.

Collision speed can play an important role in massive core formation in CCCs  \citep{2014ApJ...792...63T,2018PASJ...70S..58T}. Hydrodynamic simulations of \citet{2014ApJ...792...63T} and \citet{2018PASJ...70S..58T} have shown that high collision speeds can suppress massive core formation in CCCs. 
In Paper I, we suggested that NTSI should be strong in high-speed collisions, and strong NTSI can suppress massive core formation. 
However, a strong enough magnetic field can suppress the NTSI. Since the duration of the collision is important for core mass evolution, as demonstrated by \citet{2014ApJ...792...63T} and \citet{2018PASJ...70S..58T}, larger-sized clouds would be favorable for massive bound core formation in high-speed CCCs. \citet{2021PASJ...73S..75E} have shown evidence that massive star formation is more active in CCCs with higher collision speeds and higher column densities
in our Galaxy and nearby galaxies by compiling many observed data. We should do a systematic study with varying collision speeds and cloud sizes as in \citet{2014ApJ...792...63T} and \citet{2018PASJ...70S..58T} for colliding clouds with magnetic fields, since we have already shown the important role of magnetic field for massive bound core formation by CCCs. 
It should be very interesting to study the effect of collision speed and cloud size on massive star formation in magnetized, colliding clouds. With this aim, we extend our previous study in Paper I to collision models with higher collision speeds as high as 40 km s$^{-1}$ and larger-sized clouds as large as 20 pc. The collision speed of 40 km $^{-1}$ is higher than typical collision speeds reported in \citet{2021PASJ...73S..75E}, but such irregular velocity is observed in molecular clouds in the Galactic central region \citep{2015PASJ...67..109T}. 
The clouds used in our simulations are rather low in the mass range of observed molecular clouds. However,  the simulations of these clouds with these higher collision speeds are sufficient for a study of these effects in magnetized, colliding clouds and comparison with observed CCCs in the low column density range in \citet{2021PASJ...73S..75E}.

The plan of this paper is as follows. In section \ref{methods}, we describe the numerical method and models. In section \ref{results}, we present our numerical results. In section \ref{discuss}, we give discussion of our results. In section \ref{summmary}, we give a summary of our study.
\section{Numerical Method and Models}\label{methods}
\subsection{Numerical Method}\label{numerical_methods}
We use the same simulation method as in Paper I, and we briefly summarize it here. We use \textsc{Enzo} that is three-dimensional MHD adaptive mesh refinement (AMR) code \citep{2014ApJS..211...19B,2019JOSS....4.1636B}. 
We assume ideal MHD in our simulations. 
The code solves the MHD equations using the MUSCL 2nd-order Runge-Kutta temporal update of the conserved variables with the Harten-Lax-van Leer (HLL) method and a piecewise linear reconstruction method (PLM). 
The hyperbolic divergence cleaning method of \citet{2002JCoPh.175..645D} is adopted to ensure the solenoidal constraint on the magnetic field.
We use same numerical methods of cooling, pressure floor, and the Alfv\'{e}n speed limiter as in Paper I. 
The minimum density in our simulations is selected as the initial density of ambient medium of 1.69$\times$10$^{-23}$ g cm$^{-3}$ (see section \ref{2_2_1section}).

\begin{figure*}
    \begin{center}
    \includegraphics[width=0.9\textwidth]{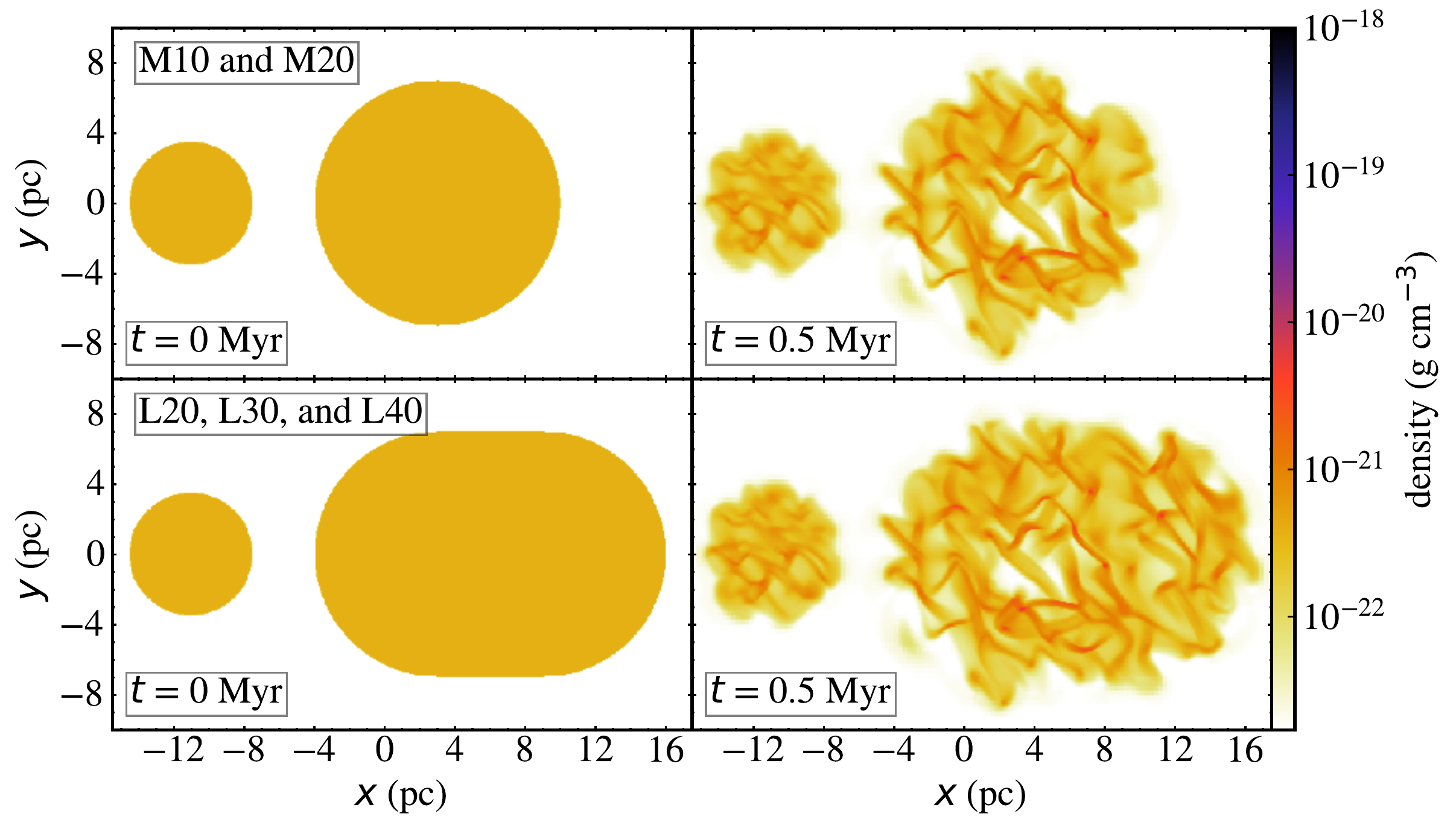}
    \end{center}
    \caption{Slice plots of the gas density in $z$ = 0 pc at $t$ = 0 (left) and 0.5 Myr (right) for M10 and M20 in top panels and L20, L30, and L40 in bottom panels. The color bar shows gas density.}
    \label{fig:dens_slice_initial} 
\end{figure*}

\begin{table*}
\caption{Initial cloud model parameters.}
\label{tab:cloud}
\centering
    \begin{threeparttable}
    \centering
    \begin{tabular}{  c  c  c  c  c  c  c  }
    \hline \hline
    Parameter & Small cloud & Medium cloud & Large cloud  & Isolated Medium cloud & Isolated Large cloud & Units \\ \hline \hline
    Shape & Sphere & Sphere  & Capsule & Sphere & Capsule & - \\ \hline
    $R$\tnote{a} & 3.5 & 7  & 7 & 7.25 & 7.18 & pc  \\ \hline
    $H$\tnote{b} & - & -  & 6 & - & 6.15 & pc  \\ \hline
    $M$\tnote{c} & 972 & 7774 & 12803 & 8566 & 13595 & $M_{\odot}$ \\ \hline
    $\rho_{0}$\tnote{d} &3.67 $\times$10$^{-22}$   & 3.67 $\times$10$^{-22}$ &  3.67 $\times$10$^{-22}$ & 3.67 $\times$10$^{-22}$ &  3.67 $\times$10$^{-22}$ & g cm$^{-3}$ \\ 
    \hline
    $t_{\textrm{ff}}$\tnote{e} & 3.5 & 3.5 & 3.5 & 3.5 & 3.5 & Myr \\ \hline
    $\sigma_{v}$\tnote{f} & 1 & 2 & 2 & 2 & 2 & km s$^{-1}$ \\ \hline \hline
    \end{tabular}
    \begin{tablenotes}
    \item[a]The radius of sphere for spherical clouds or hemispheres in capsule-shaped clouds.
    \item[b]The height of cylindrical part in capsule-shaped clouds.
    \item[c]The cloud mass.
    \item[d]The cloud initial density.
    \item[e]The free-fall time of the cloud.
    \item[f]The velocity dispersion of the cloud.
    \end{tablenotes}
    \end{threeparttable}
\end{table*}

\subsection{Cloud Models}\label{2_2section}
\subsubsection{Initial clouds and collision setup}\label{2_2_1section}
We assume two combinations of colliding clouds, which are denoted Small and Medium clouds (referred with M), and Small and Large clouds (referred with L), as shown in Figure  \ref{fig:dens_slice_initial}. We assume an initial state of each cloud based on properties of observed GMCs \citep{2009ApJ...699.1092H,2011ApJ...729..133M}, as summarized in Table 1. 

Small cloud, Medium cloud, and Large cloud have same uniform density, $\rho_0$ = 3.67 $\times $10$^{-22}$ g cm$^{-3}$, of which free-fall time is 3.5 Myr. Small and Medium clouds are spherical clouds with radii of 3.5 pc and 7 pc, respectively, as in Paper I. 
Masses of Small and Medium  clouds are 972 $M_{\odot}$ and 7774 $M_{\odot}$, respectively. 
We assume Large cloud with a larger size along the collision axis ($x$-axis of the simulation box) to study effect of long duration of collision between the clouds on massive core formation. 
Large cloud initially has an elongated shape (capsule) consisting of a cylinder with two hemispheres at its both ends. 
The height of this cylinder is 6 pc, and the radii of the two hemispheres are 7 pc. Its symmetric axis is along the collision axis.
Mass of Large cloud is 12803 $M_{\odot}$. 
The lengths of Medium and Large clouds along the collision axis are 14 pc and 20 pc, respectively.  

We assume initial temperatures of the clouds as 68 K, 273 K, and 273 K for Small cloud, Medium cloud, and Large cloud, respectively. These temperatures provide the initial thermal pressure support for the clouds. The dense gas in the clouds rapidly cools down to 10 K due to the radiative cooling during their evolution. These parameters of the clouds are summarized in Table \ref{tab:cloud}.
The ambient medium has a density of 1.69 $\times$ 10$^{-23}$ g cm$^{-3}$ and a temperature of 800 K.
This high density of the ambient medium is used to avoid high Alfv\'{e}n speeds in the ambient medium.

These clouds are immersed in an initial uniform magnetic field, $\boldsymbol B_0$, and turbulence is initially generated in both clouds (see Section \ref{magnetic-field}). 
After $t$ = 0.5 Myr for their isolated evolution, collision speed is given to Small cloud in the direction of other cloud. 
We call the other cloud the target cloud.

We simulated five CCC models, as shown in Table \ref{tab:model}. Two additional isolated cloud models are simulated for comparison. For the collision of Small and Medium clouds, we assume collision speeds of 10 and 20 km s$^{-1}$ (M10 and M20). For the collision of Small and Large clouds, collision speeds of 20, 30, and 40 km s$^{-1}$ are assumed (L20, L30, and L40).
For isolated clouds, Isolated Medium cloud in IM0 model is a spherical cloud with a total mass of Small and Medium clouds combined, and Isolated Large cloud in IL0 model is an elongated cloud with similar shape of Large cloud with a total mass of Small and Large clouds combined. Since we want to make clear the effects of collision by comparison  with numerical results of the isolated clouds, we assume each isolated cloud with same total mass of corresponding, colliding clouds. For this purpose, we assume same initial density $\rho_0$ for isolated clouds. Detailed parameters of these isolated clouds are given in Table \ref{tab:cloud}.
We stop our simulations at $t$ = 3.1 Myr, which is earlier than the free-fall time of the clouds (3.5 Myr).

Our simulation domain encompasses (64 pc)$^3$ with root grids 256$^3$, and we use four refinement levels based on the condition of minimum baryon mass of 0.05 $M_{\odot}$ for refinement. This gives the minimum cell size of 0.015 pc at the maximum refinement level.

The details of turbulence and magnetic fields are mentioned in Section \ref{magnetic-field}. The methods used for the analysis of dense cores are mentioned in Section \ref{model-densecore}.

\begin{table}
\caption{Simulation models.}
\label{tab:model}
\centering
    \begin{threeparttable}
    \centering
    \begin{tabular}{p{1cm} p{1cm} p{1cm} p{1cm} p{1.5cm}}
    \hline \hline
    Model & ${B_0}$ (\textmu G)\tnote{a} & \multicolumn{2}{|c|}{Clouds} & $v_{\textrm{coll}}$ (km s$^{-1}$)\tnote{b} \\
    Name& & Left & Right 
    \\ \hline \hline 
    M10\tnote{c} & 4.0 & Small & Medium & 10  \\ \hline
    M20 & 4.0 & Small & Medium & 20  \\ \hline
    L20 & 4.0 & Small & Large & 20  \\ \hline
    L30 & 4.0 & Small & Large & 30  \\ \hline
    L40 & 4.0 & Small & Large & 40  \\ \hline
    IM0 & 4.0 & - & Isolated Medium & -  \\ \hline
    IL0 & 4.0 & - & Isolated Large & -  \\ \hline \hline
    \end{tabular}
    \begin{tablenotes}
    \item[a]The initial magnetic field strength. 
    \item[b]The collision speed given to the left cloud.
    \item[c]Same as the Ystrong model in Paper I, expect some differences in initial velocity field mentioned in main text.
    \end{tablenotes}
    \end{threeparttable}
\end{table}

\subsubsection{Magnetic field and Turbulence in clouds}\label{magnetic-field}
After the clouds are initially immersed in a uniform magnetic field, $\boldsymbol B_0$, we develop turbulent motions inside them from $t$ = 0 to 0.5 Myr, resulting in turbulent magnetic fields inside the clouds.
The strength of initial magnetic field $\boldsymbol B_0$ is assumed to be $B_0$ = 4.0 \textmu G, as in Paper I, since we have shown that turbulent magnetic fields in the turbulent clouds in this choice are consistent with the observed relation between gas densities and turbulent magnetic fields in molecular clouds given by \citet{2010ApJ...725..466C}. The direction of $\boldsymbol B_0$ is normal (the positive $y$-axis of the simulation box) to the collision axis. We use this direction of $\boldsymbol B_0$ in all our simulations. This is because we found that the total number of dense cores more than 10 $M_{\odot}$ is not so different between models with different directions of $\boldsymbol B_0$ in Paper I, although detailed evolution of density structures and dense cores is different between them.  

Turbulent velocities are generated to be consistent with the Larson relation \citep{1981MNRAS.194..809L, 2009ApJ...699.1092H} at $t$ = 0 Myr, by imposing a velocity field with power spectrum ${v_k}^{2}$ $\propto$ $k^{-4}$. 
We assume the velocity dispersion, $\sigma_v$ $\sim$ 1.0 km s$^{-1}$ for Small cloud and $\sigma_v$ $\sim$ 2.0 km s$^{-1}$ for Medium and Large clouds. We use the same box which covers the Large cloud and generate turbulent velocity fields. We adapt velocity fields in the target cloud. By this choice we have similar density structures in 
overlap regions in Medium and Large clouds at $t$ = 0.5 Myr, as shown in right panels of Figure \ref{fig:dens_slice_initial}. Since we use different size of the box from Paper I for the turbulence generation, the detailed density structures in the target clouds at $t$ = 0.5 Myr are different from Paper I at $t$ = 0.5 Myr.

\subsubsection{Dense cores}\label{model-densecore}

In order to study dense core formation and evolution, we define a dense core by a threshold density, $\rho_{\textrm{th}}$ = 5 $\times$ 10$^{-20}$ g cm$^{-3}$ ($\sim$ 136$\rho_0$), which is in the range of the typical density of observed molecular cores \citep{2007ARA&A..45..339B}, as in Paper I. 
We define dense cores by following steps: 1) selection of cells with $\rho$ $\ge$ $\rho_{\textrm{th}}$ as dense cells, 2) grouping together the neighboring dense cells into a dense core, and 3) exclusion of those groups with cell number less than 27 from dense cores. This minimum cell number condition is used to get a good spatial resolution of dense cores.

Self-gravitationally bound dense cores are expected to form stars, and self-gravitationally bound dense cores greater than 10 $M_{\odot}$ are expected to form massive stars. We check the gravitational boundness of each dense core by comparing its turbulent energy, $E_{\textrm{turb}}$, its magnetic field energy, $E_{\textrm{mag}}$, its thermal energy, $E_{\textrm{ther}}$, and its self-gravitational energy, $E_{\textrm{grav}}$. 
$E_{\textrm{turb}}$ is given for a dense core by
\begin{equation}\label{eq:eturb}
E_{\textrm{turb}}=\sum_i \frac{1}{2}m_i|\boldsymbol v_i-\boldsymbol v_{\textrm{mean}}|^2,
\end{equation}
where $i$ is an index of a dense cell in the dense core, the sum is made over all cells in the dense core, $m_i$ is the mass of the dense cell $i$, $\boldsymbol v_i$ is the velocity of the dense cell $i$, and $\boldsymbol v_{\textrm{mean}}$ is the mean velocity of the dense core given by
\begin{equation}
\boldsymbol v_{\textrm{mean}}= \frac{\sum_im_i\boldsymbol v_i}{\sum_im_i}.
\end{equation}
$E_{\textrm{mag}}$ is given by 
\begin{equation}\label{eq:emag}
E_{\textrm{mag}}=\sum_i \frac{\boldsymbol B_i \cdot \boldsymbol B_i}{8\pi }V_i,
\end{equation}
where $\boldsymbol B_i$ and $V_i$ are the magnetic field flux density vector and volume of the dense cell $i$, respectively.
$E_{\textrm{ther}}$ is given by 
\begin{equation}\label{eq:ether}
E_{\textrm{ther}}=\sum_i \frac{3}{2}m_i{c{_{s, i}}}^2,
\end{equation}
where $c_{s, i}$ is the sound speed of the dense cell $i$.
We estimate $E_{\textrm{grav}}$ by
\begin{equation}\label{eq:egrav}
E_{\textrm{grav}} =- \frac{3GM_{\textrm{core}}^2}{5\ensuremath{\langle R \rangle}},
\end{equation}
where $G$ is the gravitational constant, $M_{\textrm{core}}$ is mass of the dense core, and $\ensuremath{\langle R \rangle}$ is given by 
\begin{equation}\label{eq:rad}
\ensuremath{\langle R \rangle} = {\left(\frac{3V_{\textrm{core}}}{4\pi}\right)^{1/3}},
\end{equation}
where $V_{\textrm{core}}$ is total volume of the dense core. 
If $(E_{\textrm{turb}}+E_{\textrm{mag}}+E_{\textrm{ther}})$ $\leq$ $|E_{\textrm{grav}}|$, the dense core is gravitationally bound, and we call such a dense core a bound core. 
Since a dense core is selected by using condition of $\rho$ $\geq$ $\rho_{\textrm{th}}$, its free-fall time $t_{\textrm{ff}}$ $\leq$ 0.3 Myr. 

To understand the time evolution of dense core and its physical properties, we trace their evolution using a similar method as in \citet{2014ApJ...792...63T}.
We modify their method as follows. We increase size of a spherical volume for searching of the next core position around its predicted position in their method by a factor of 1.75 (see Section 2.1 in their paper for details). We use additional condition that mass of the core in the next time step should be greater than 0.2 times its mass at the previous time step of data-output of which interval is 0.1 Myr. By these modifications, we exclude very small core which is most close to the predicted position, since we find more massive core in the increased spherical volume in some model.

To make clear the role of accretion in the mass evolution of dense cores, we estimate accreted mass onto dense cores based on Bondi accretion \citep{1952MNRAS.112..195B}. Following the core evolution, we estimate the mass due to accretion up to the ($n$+1)$^{\textrm{th}}$ simulation time step as
\begin{equation}\label{eq:accretedmass}
M_{\textrm{acc}} = M_{\textrm{init, core}} + \sum_{n=1}^{n}\dot{M_n}{\Delta t},
\end{equation}
where $M_{\textrm{init, core}}$ is core mass at formation epoch of the dense core ($n$ = 1), $\Delta t$ is the time interval between the simulation time steps, and $\dot{M_n}$ is accretion rate of the core at $n^{\textrm{th}}$ time step defined by
\begin{equation}\label{eq:accetion}
\dot{M_n} = \pi{r^2_{\textrm{acc}}}\sigma_{\textrm{srr}}\rho_{\textrm{srr}},
\end{equation}
where $\sigma_{\textrm{srr}}$ is the average effective speed given by 
\begin{equation}\label{eq:accetion2}
\sigma_{\textrm{srr}} = (c^2_{\textrm{s, srr}} + \sigma^2_{\textrm{1D, srr}} + v^2_{\textrm{A, srr}})^{1/2},
\end{equation}
where $c_{\text{s, srr}}$, $\sigma_{\textrm{1D, srr}}$, $v_{\textrm{A, srr}}$, and $\rho_{\textrm{srr}}$ are the mass-weighted averages of sound speed, 1D non-thermal velocity dispersion, Alfv\'{e}n speed, and density in a volume of a sphere of radius $r_{\textrm{srr}}$ around the core excluding the core volume, respectively. The $r_{\textrm{srr}}$ is calculated from gas properties in a volume surrounding the core. We repeat the following procedure to calculate $r_{\textrm{srr}}$ and $r_{\textrm{acc}}$. 
We use the formula,
\begin{equation}
r_{\textrm{sph}} = \frac{2GM_{\textrm{core}}}{\sigma^2} + \ensuremath{\langle R \rangle},
\end{equation}
which is sum of the Bondi radius and $\ensuremath{\langle R \rangle}$ given by equation (\ref{eq:rad}).
First, we calculate $r_{\textrm{sph}}$ by this equation using  $\sigma^2$ in the core. Then, we obtain the mean value of $\sigma^2$ in a volume of a sphere of $r_{\textrm{sph}}$ outside of the core.
Second, we obtain again $r_{\textrm{sph}}$  using the new $\sigma^2$. We use the new $r_{\textrm{sph}}$ as $r_{\textrm{srr}}$. Finally, after we calculate the mean value of $\sigma^2$ in a volume of a sphere of $r_{\textrm{srr}}$ outside of the core, we obtain $r_{\textrm{sph}}$ using the second new $\sigma^2$. We use this $r_{\textrm{sph}}$ as $r_{\textrm{acc}}$.
In this procedure, we assume that maximum radius of these spheres is 1 pc.
 
We estimate power index of core mass function for cores greater than 10 $M_{\odot}$ as follows. If the core mass function, $\phi$, is defined as
\begin{equation}\label{phi-eq}
\phi = \frac{dN}{dM_{\rm core}} \propto M_{\rm core}^{-\gamma}.
\end{equation}
where $dN$ is the number of cores between $M_{\rm core}$ and $M_{\rm core}$+$dM_{\rm core}$, and $\gamma$ is a power index of the core mass function, the cumulative core mass distribution, $N(\geq M_{\rm core})$, is given by
\begin{equation}
N(\geq M_{\rm core})=\int_{M_{\rm core}}^{\infty} \frac{dN}{dM_{\rm core}}dM_{\rm core} \propto M_{\rm core}^{-(\gamma-1)}\propto M_{\rm core}^{\alpha}, \label{cmf_fit}
\end{equation}
where $\alpha$ = -($\gamma$-1). 
We fit equation (\ref{cmf_fit}) to the cumulative core mass distribution of dense cores obtained in our numerical results to get the power index of the core mass function, $\gamma$.
\section{Numerical Results}\label{results}
We present simulation results of collision of Small and Medium clouds (M10 and M20) in Section \ref{coll_sp_effect}, and those of \textcolor{red}{the} collision of Small and Large clouds (L20, L30, and L40) in Section \ref{cld_size_effect}.

\subsection{Collision of Small and Medium clouds}\label{coll_sp_effect}
\begin{figure*}
    \begin{center}
    \includegraphics[width=0.9\textwidth]{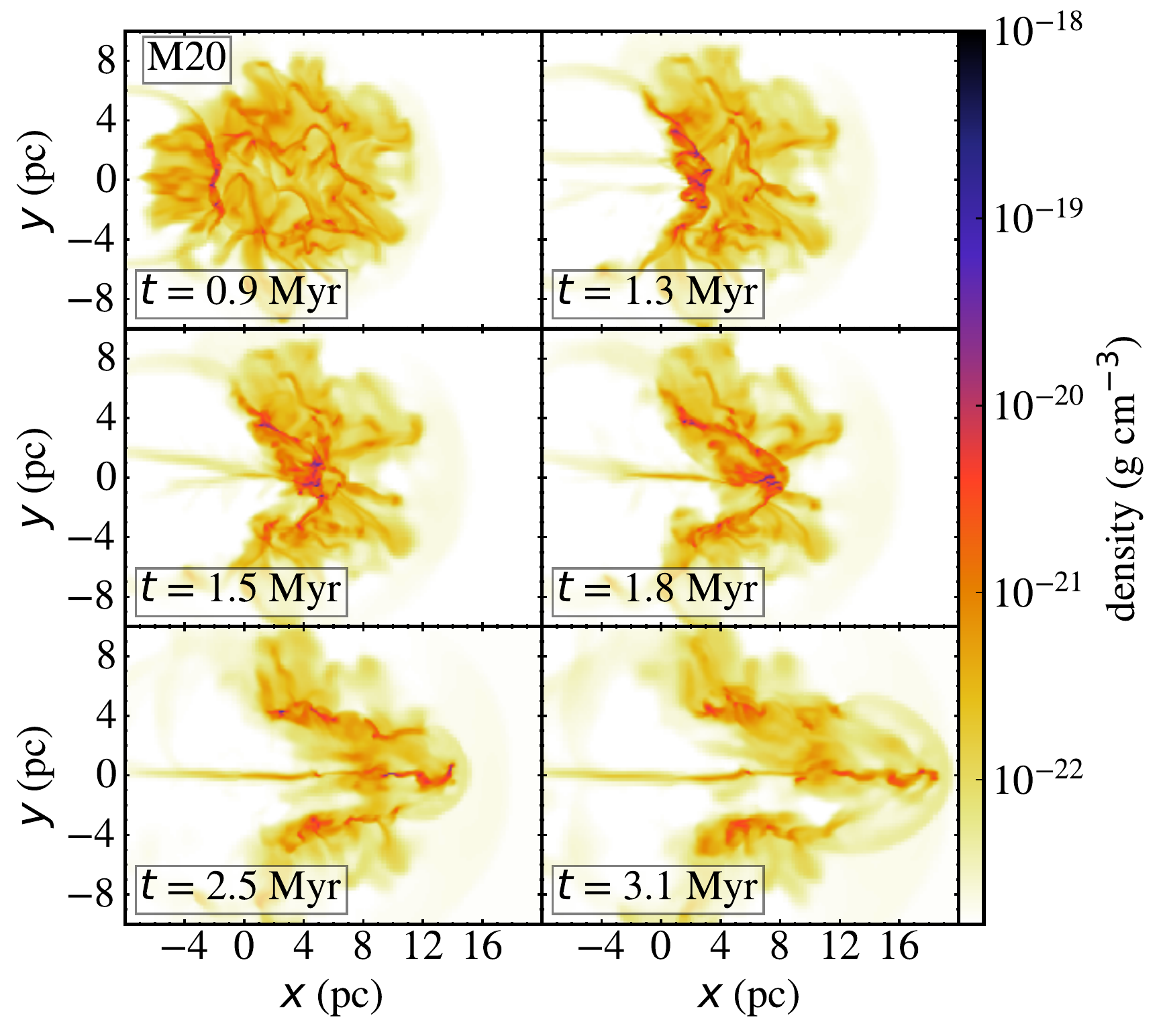}
    \end{center}
\caption{Slice plots of the gas density in $z$ = 0 pc at $t$ = 0.9, 1.3, 1.5, 1.8, 2.5, and 3.1 Myr in M20. The color bar shows gas density.}
\label{fig:dens_slice_M20strong} 
\end{figure*}

\begin{figure*}
 \begin{tabular}{cc}
 \begin{minipage}[t]{0.5\hsize}
    \begin{center}
    \includegraphics[width=\textwidth]{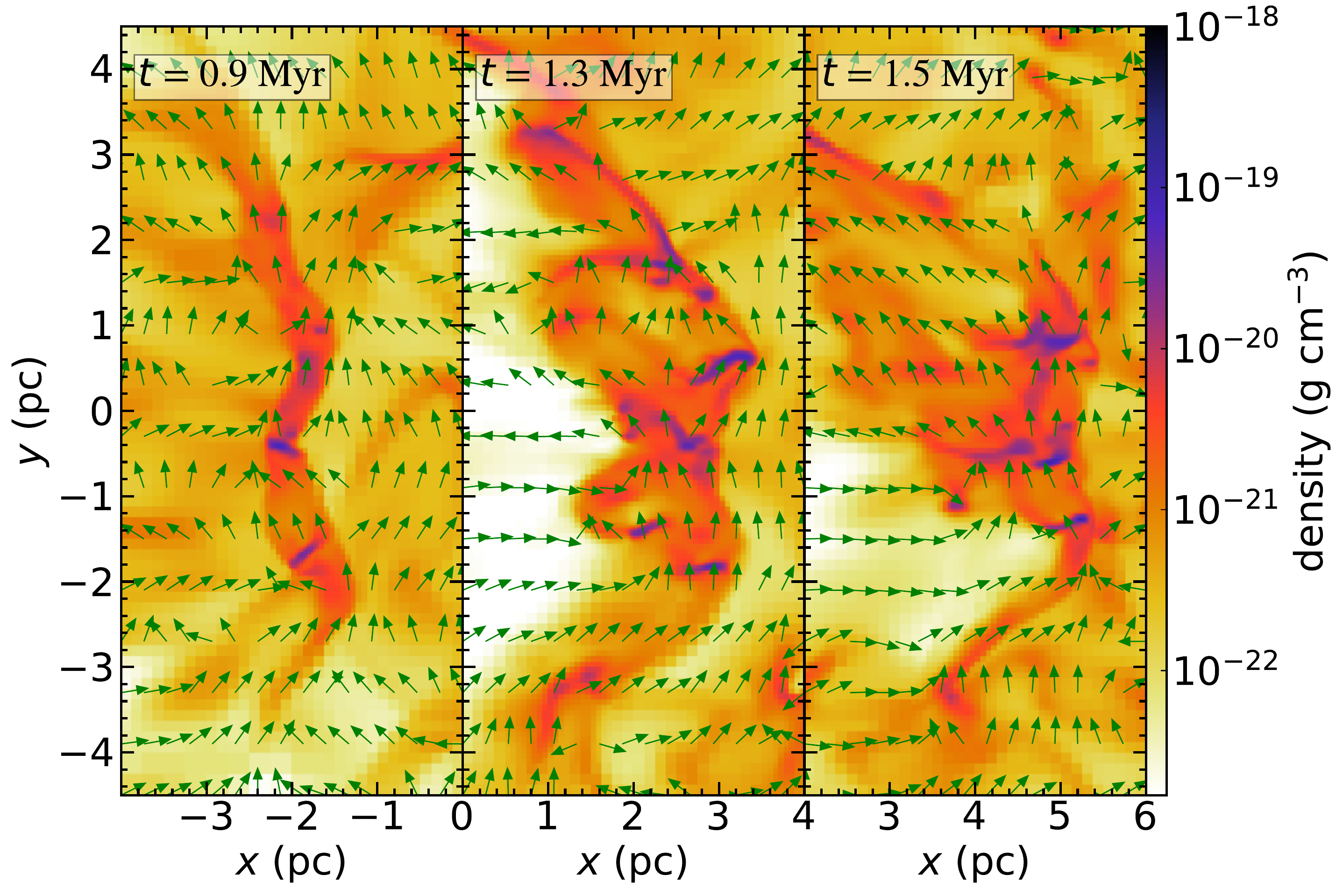}
    \end{center}
    \end{minipage}
    \begin{minipage}[t]{0.5\hsize}
    \begin{center}
    \includegraphics[width=\textwidth]{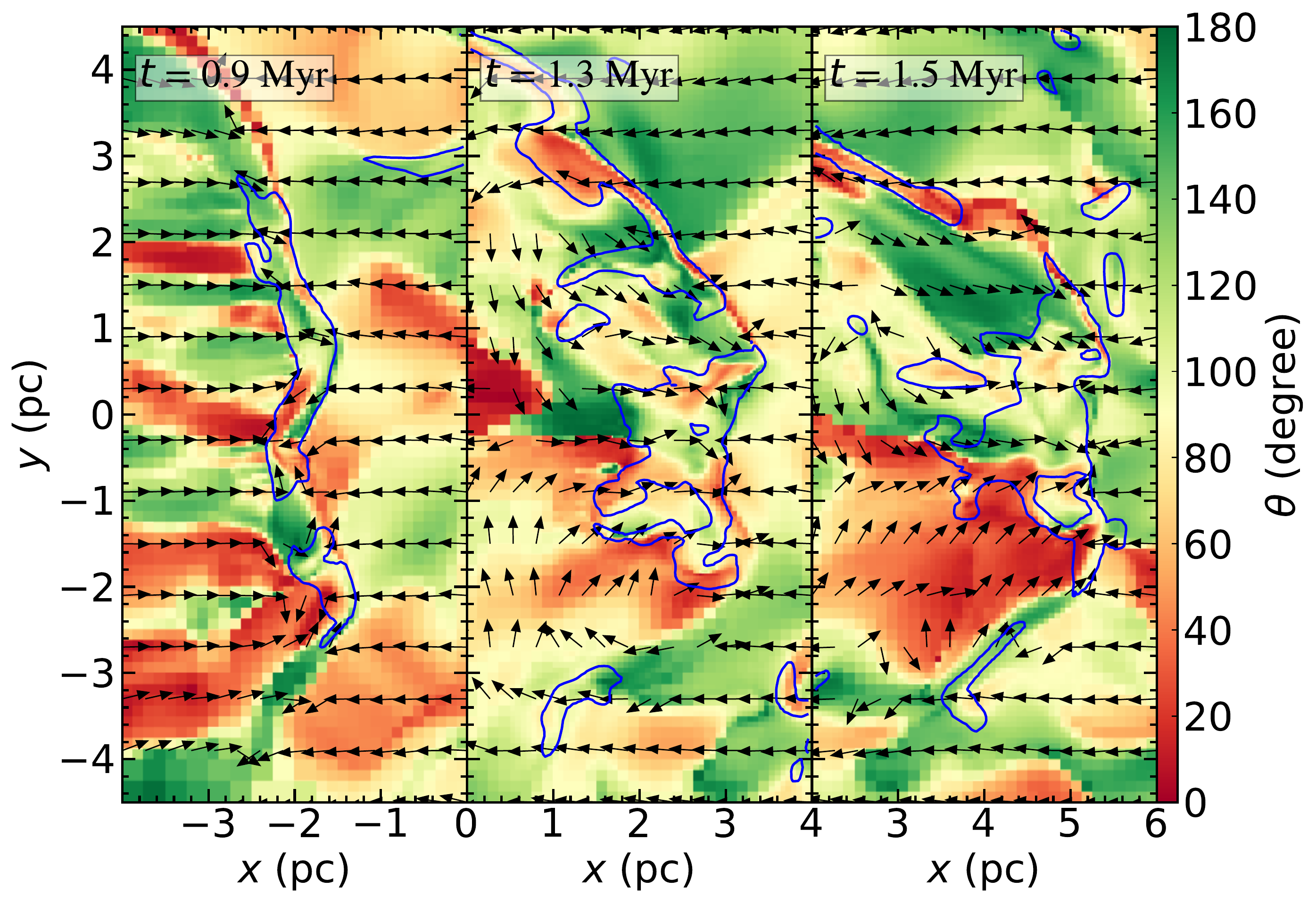}
    \end{center}
    \end{minipage}
 \end{tabular}
\caption{Left panels: Slice plots of the gas density in $z$ = 0 pc at $t$ = 0.9, 1.3, and 1.5  Myr with normalized vectors of magnetic filed directions given by $(B_x, B_y)/\sqrt{B_x^2+B_y^2}$ in M20. Right panels: Slice plots of $\theta$ which is the angle between the magnetic field and the velocity ($v_x, v_y, v_z$), in $z$ = 0 pc at same epochs with unit vectors of $(v_x, v_y)/\sqrt{v_x^2+v_y^2}$ in M20. The $v_x$, $v_y$, and $v_z$ are $x$-, $y$-, and $z$-components of velocity in the frame co-moving with the shocked region, respectively. The contours in right panels show gas with a density of more than 5$\rho_0$.}
\label{fig:bend} 
\end{figure*}

This section shows simulation results of the collision of Small and Medium clouds. We show results of 20 km s$^{-1}$ collision (M20) in Section \ref{section:M20}. 
In Section \ref{section:M10}, we compare M20 with 10 km s$^{-1}$ collision (M10) results. In Section \ref{section:M_densecore}, we compare dense core formation and evolution in both models.
\subsubsection{Time evolution of M20}\label{section:M20}
Figure \ref{fig:dens_slice_M20strong} shows time evolution of the gas density structures of M20 in which the collision speed is 20 km s$^{-1}$. The top and middle panels in Figure \ref{fig:dens_slice_M20strong} show early stage of the collision at $t$ = 0.9 Myr, the shocked region having crossed Small cloud at $t$ = 1.3 Myr, the shocked region proceeding in Medium cloud at $t$ = 1.5 Myr, and the shocked region near Medium cloud's right edge at $t$ = 1.8 Myr.

At $t$ = 0.9 Myr, the shocked region formed by the supersonic collision of both clouds is near $x$ = -2 pc and has two shock fronts (left and right shock fronts) on both sides. NTSI on scales smaller than 1 pc is well suppressed by the magnetic fields, although the collision speed is twice of Paper I in which we simulated a collision of the same clouds with the collision speed of 10 km s$^{-1}$ and with the same initial magnetic field in M20. 
In Paper I, such spatial shifts are suppressed for $B_0$ = 4.0 \textmu G and develop for $B_0$ = 0.1 \textmu G. Dense structures more than the core threshold density, 5 $\times$ 10$^{-20}$ g cm$^{-3}$, (hereafter, we call these structures as the high-density gas regions) are formed in the shocked region.

\begin{figure*}
    \begin{center}
    \includegraphics[width=1\textwidth]{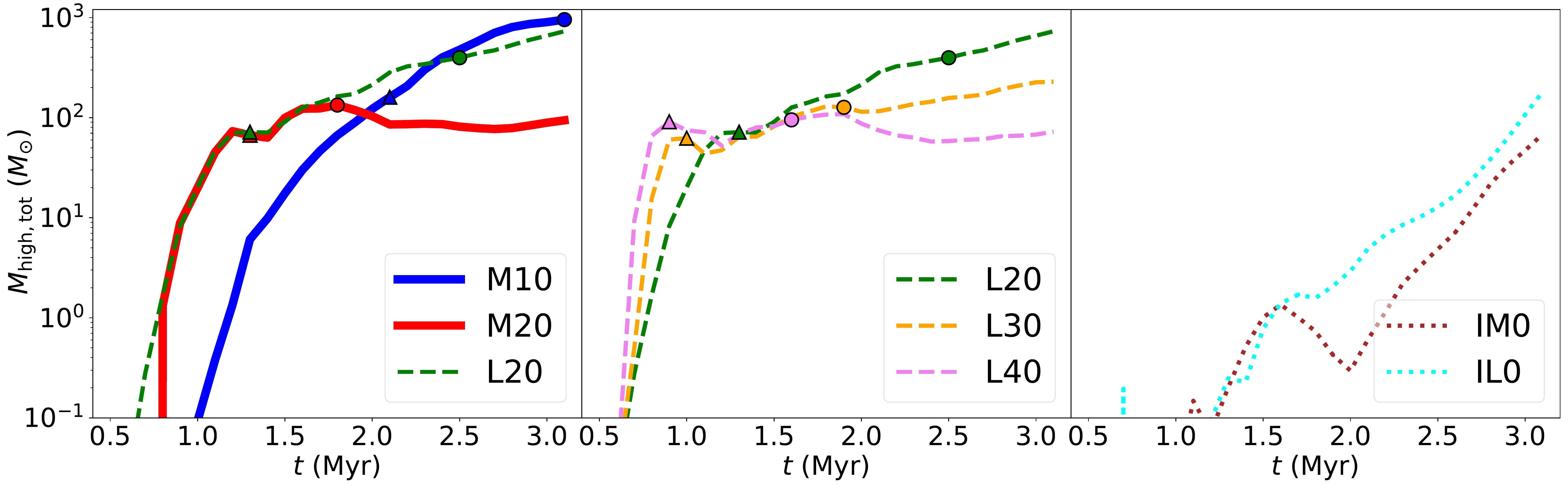}
    \end{center}
    \caption{Time evolution of total mass of high-density (> 5 $\times$ 10$^{-20}$ g cm$^{-3}$) gas regions, $M_{\mathrm{high,tot}}$, in all models. The triangles and circles indicate the shock-crossed epochs of Small cloud and target clouds for all CCC models, respectively. The triangles for M20 and L20 are overlapping in the early phase.}
    \label{fig:highdenseCCC}
\end{figure*} 

\begin{figure*}
    \begin{center}
    \includegraphics[width=0.7\textwidth]{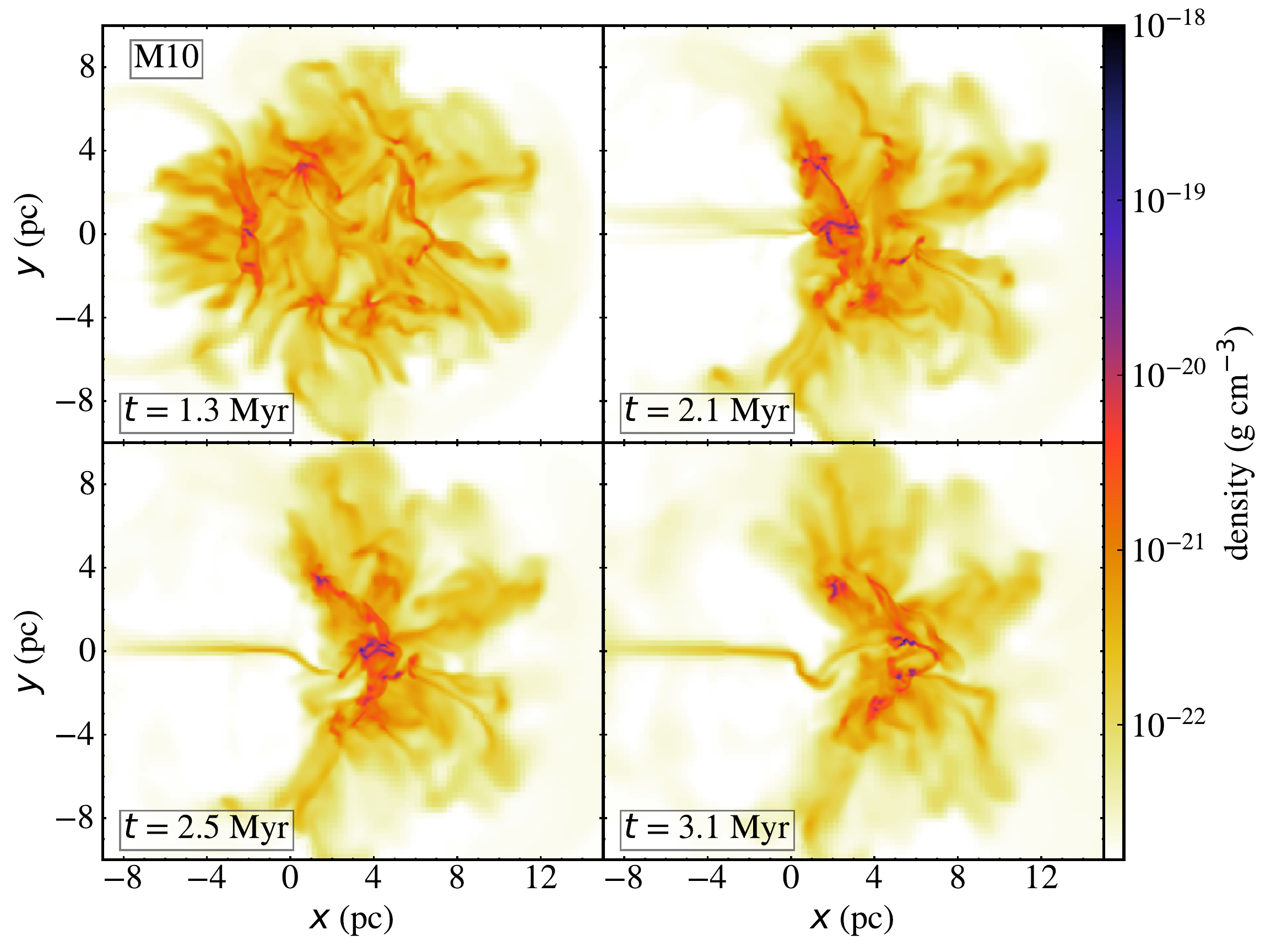}
    \end{center}
    \caption{Slice plots of the gas density in $z$ = 0 pc at $t$ = 1.3, 2.1, 2.5, and 3.1 Myr in M10. The color bar shows gas density.}
    \label{fig:dens_slice_M10strong}
\end{figure*} 

At $t$ = 1.3 Myr, Small cloud has completely entered the shocked region leaving a low-density cavity on the left-hand side of the shocked region. Hereafter, we call this epoch the shock-crossed epoch of Small cloud.
Several high-density gas regions are already formed in the shocked region. The shape of the shocked region is roughly arc-like due to the difference in sizes of the colliding clouds. The thickness of the shocked region is still increasing with time after $t$ = 1.3 Myr.

At $t$ = 1.5 Myr, the left-hand side edge of the shocked region is relatively expanding to the negative direction of the $x$-axis. 
This is because the ram pressure by Small cloud to the left side of the shocked region disappears after the shock-crossed epoch of Small cloud. 
Most of the high-density gas regions in the shocked region move toward the collision axis ($y$ = 0 pc). The arc-like structure of the shocked region is more bent than that at $t$ = 1.3 Myr.

At $t$ = 1.8 Myr, the shocked region reaches the right edge of Medium cloud. Hereafter, we call this epoch the shock-crossed epoch of Medium cloud. A large high-density gas region is formed near the collision axis at this epoch, caused by the accumulation of dense gas regions by gas flow along the magnetic field in the arc-like shocked region, as shown in Figure \ref{fig:bend}. 

We show close-up views of magnetic field structures on colormaps of the density in the shocked region at $t$ = 0.9, 1.3, and 1.5 Myr
in the left panels in Figure \ref{fig:bend}. We show magnetic field directions with unit vectors defined by $(B_x, B_y)/\sqrt{B_x^2+B_y^2}$. These panels show that the magnetic fields in the shocked region roughly align with the edge of the shocked region. We show gas flow structures with unit vectors defined by $(v_x, v_y)/\sqrt{v_x^2+v_y^2}$
in the right panels in Figure \ref{fig:bend}.
We also show colormaps of $\theta$ which is an angle between magnetic field and velocity ($v_x$, $v_y$, $v_z$) in the right panels to make clear relative directions between them, where $v_x$, $v_y$, and $v_z$ are 3D-components of gas velocity in the frame co-moving with the shocked region. The contours show gas with a density of more than 5$\rho_0$ which roughly indicates the shocked region. The right panels show that the gas flow in upper and lower regions of the shocked region is roughly along the magnetic fields at $t$ = 1.3 Myr and more at $t$ = 1.5 Myr, as indicated by $\theta$ $\sim$ 180 or 0 degrees. Figure \ref{fig:bend} indicates that the magnetic fields in the shocked region bend along the deformation of the shocked region, gas is easy to move along the deformed magnetic field lines in the shocked region, and this gas flow helps the accumulation of the high-density gas regions to the collision axis. 

The gas flow along the deformed magnetic field lines can be driven by inertial force due to the deceleration of the shocked region. This is estimated as follows. The mass-weighted velocity of the shocked region decelerates with time from 8.6 km s$^{-1}$ at $t$ = 1.3 Myr to 6.5 km s$^{-1}$ at $t$ = 1.5 Myr.  
From $t$ = 1.3 to 1.5 Myr, this deceleration can drive gas flow along the magnetic field lines, since the magnetic fields in the shocked region are large enough to control the gas flow along the magnetic fields. This flow speed can be estimated as
\begin{equation}
v = (8.6-6.5)\cos\theta \textrm{ km s}^{-1} =  1.4\cos\theta/\cos(\pi /4) \textrm{ km s}^{-1},
\end{equation}
where $\theta$ is an angle between the deceleration and the magnetic field.
This flow speed is high enough to explain the gas flow that contributes to the accumulation of the high-density gas regions to the collision axis.

The bottom panels in Figure \ref{fig:dens_slice_M20strong} show the time evolution of the gas density structures after the shock-crossed epoch of Medium cloud. These panels show that the gas in the leading part of the shocked region expands in the ambient medium. At $t$ = 2.5 Myr, the gas expansion already occurs.
This expansion begins near the shock-crossed epoch of Medium cloud and is due to excess magnetic pressure in the shocked region than ram pressure by the ambient medium after the shock-crossed epoch of Medium cloud.
The mean magnetic pressure in the shocked region is $\sim$ 5 $\times$ 10$^{-11}$ g cm$^{-1}$ s$^{-2}$ at the shock-crossed epoch of Medium cloud, which is significantly higher than that in the right-hand side of the shocked region.
Before the shock-crossed epoch of Medium cloud, the shocked region is confined by the ram pressure by Medium cloud, $\sim$ $\rho_{\textrm{0}}{v_{\textrm{sh}}}^2$ $\sim$ 7 $\times$ 10$^{-11}$ g cm$^{-1}$ s$^{-2}$, which is comparable to the magnetic pressure of the shocked region. 
After the shock-crossed epoch of Medium cloud, the ram pressure is provided by the ambient medium and is significantly lower than the ram pressure by Medium cloud.
This leads to the expansion of the leading part of the shocked region. Details of this expansion and its effects on the dense cores are given in Section \ref{section:M_densecore}.

Figure \ref{fig:highdenseCCC} shows the time evolution of the total mass of the high-density regions, $M_{\mathrm{high,tot}}$, in M20 (red thick-line in left panel) and other models. $M_{\mathrm{high,tot}}$ increases with time in M20 until $t$ = 1.8 Myr (also marked by circles). After this, the changes in $M_{\mathrm{high,tot}}$ are small. In this phase, the leading part of the shocked region expands. This indicates that change of the physical state of the  shocked region affects evolution of $M_{\mathrm{high,tot}}$.

\begin{figure*}
    \begin{center}
    \includegraphics[width=1\textwidth]{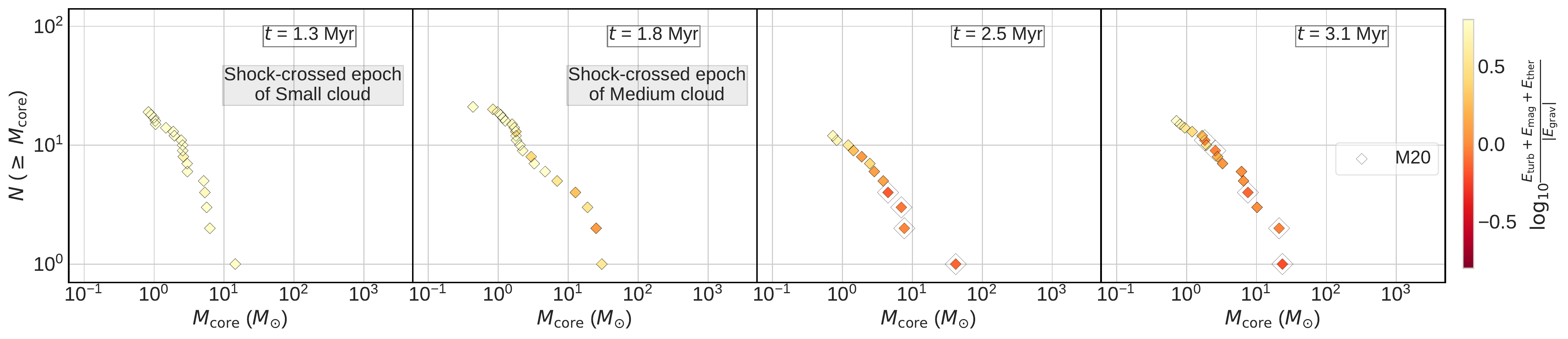}
    \end{center}
    \caption{Cumulative dense core mass distributions shown by filled diamonds at $t$ = 1.3, 1.8, 2.5, and 3.1 Myr for M20. The color bar in the right-hand side shows the energy ratio of sum of turbulent, magnetic field, and thermal energies to absolute value of self-gravitational energy. The gravitationally bound cores are marked by larger, open diamonds.}
    \label{fig:cmf_M20strong} 
\end{figure*}
\begin{figure*}
    \begin{center}
    \includegraphics[width=1\textwidth]{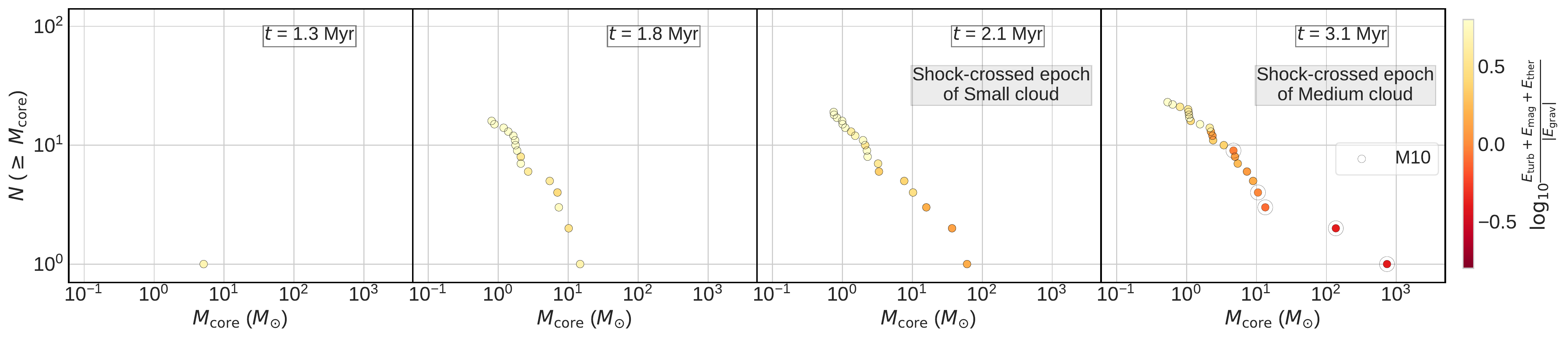}
    \end{center}
    \caption{Same as in Figure \ref{fig:cmf_M20strong} but for M10 shown by filled circles at $t$ = 1.3, 1.8, 2.1, and 3.1 Myr.}
    \label{fig:cmf_M10strong} 
\end{figure*}
\subsubsection{Comparison of time evolution of M20 and M10}\label{section:M10}
Figure \ref{fig:dens_slice_M10strong} shows the time evolution of the gas density structures of M10 in which the collision speed is 10 km s$^{-1}$. 
Each panel shows the shocked region corresponding to the top and middle panels of Figure \ref{fig:dens_slice_M20strong} which show numerical results of M20. These figures show that the time evolution of the gas structure is much faster in M20 than that in M10. The swept-up mass by the shock should be similar for the same corresponding positions of the shocked regions in the colliding clouds. However, $M_{\mathrm{high,tot}}$ evolves differently in these models as shown by thick-lines in left panel of Figure \ref{fig:highdenseCCC}.
We notice that $M_{\mathrm{high,tot}}$ is smaller than the swept-up mass. The shock-crossed epochs of Small cloud are shown by triangles in Figure \ref{fig:highdenseCCC}. 
At these epochs, the swept mass by the shock is estimated to be twice of Small cloud mass, $\sim$ 2000 $M_{\odot}$.
However, $M_{\mathrm{high,tot}}$ is much smaller than this value in all CCC models in Figure \ref{fig:highdenseCCC}. These results indicate that formation of high-density gas regions proceeds more slowly than the mass growth of the shocked region.

In the early stage of the collision (top left panels in Figure \ref{fig:dens_slice_M20strong} and Figure \ref{fig:dens_slice_M10strong}), the shocked region is thinner in M20 than that in M10. 
This is because of higher ram pressure of the pre-shock gas flows in M20 than M10. 
At $t$ = 1.3 Myr, a greater number of the high-density gas regions are formed in M20 than M10. This is because of higher mass of the shocked region in M20 due to the higher collision speed of clouds than M10 at this epoch.

At the stage of the shock-crossed epoch of Small cloud (top right panels in Figure \ref{fig:dens_slice_M20strong} and Figure \ref{fig:dens_slice_M10strong}), the high-density gas regions are less concentrated towards the collision axis in M20 than in M10.
This is because M20 has less time for their concentration to the collision axis than M10.
The concentration is driven by the flow along the deformed magnetic fields.
A similar difference appears in middle left panel in Figure \ref{fig:dens_slice_M20strong} and bottom left panel in Figure \ref{fig:dens_slice_M10strong}.
 
At the stage of the shock-crossed epoch of Medium cloud (middle right panel in Figure \ref{fig:dens_slice_M20strong} and bottom right panel in Figure \ref{fig:dens_slice_M10strong}), the high-density gas regions are highly concentrated to the collision axis in both models. However, $M_{\mathrm{high,tot}}$ is significantly lower in M20 than M10 at this stage, as shown in Figure \ref{fig:highdenseCCC}.

\begin{figure}
    \begin{center}
    \includegraphics[width=0.5\textwidth]{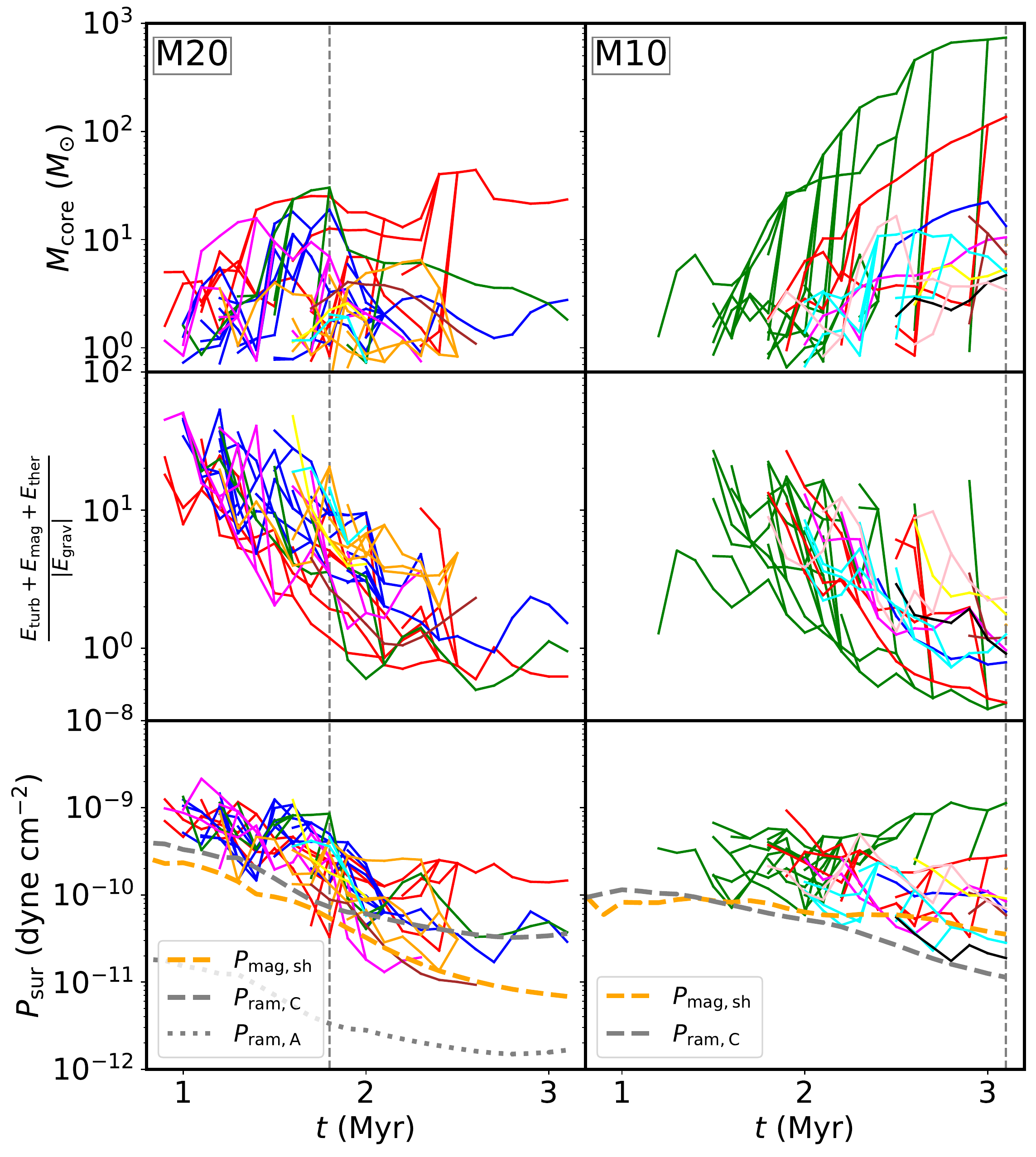}
    \end{center}
 
\caption{Time evolution of masses, the gravitational boundness ratios, and the estimated pressures at the surfaces of massive dense cores which are top ten at the shock-crossed epoch of Medium cloud, where this epoch is $t$ = 1.8 Myr in M20 (left panels) and  $t$ = 3.1 Myr in M10 (right panels). The vertical line shows the shock-crossed epoch of Medium cloud in each model. The details of other thick lines in bottom panels are given in the main text.}
\label{fig:energies2} 
\end{figure}

$M_{\mathrm{high,tot}}$ is much less in M20 than that in M10 (blue thick-line) at the end of the simulation, $t$ = 3.1 Myr, as shown in Figure \ref{fig:highdenseCCC}. $M_{\mathrm{high,tot}}$ increases with time until $t$ = 1.8 Myr and decreases after this in M20. This mass decrease occurs due to the expansion of the shocked region after $t$ = 1.8 Myr in M20, as shown in section \ref{section:M20}.
In M10, such expansion of the shocked region does not occur before $t$ = 3.1 Myr.
We will discuss the reason for the difference in evolution of $M_{\mathrm{high,tot}}$ between two models by comparing dense core evolution in both models in the next subsection.

\subsubsection{Dense core formation and evolution in M20 and comparison with M10}\label{section:M_densecore}

We show cumulative dense core mass distributions (CMDs), $N$ ($\ge$ $M_{\textrm{core}}$), which is the number of dense cores with a mass more than $M_{\textrm{core}}$ as given in equation (\ref{cmf_fit}), at $t$ = 1.3, 1.8, 2.5, and 3.1 Myr for M20 in Figure \ref{fig:cmf_M20strong}.
In this figure, the color of filled diamond markers shows the logarithmic value of gravitational boundness ratio defined as (($E_{\textrm{turb}}+E_{\textrm{mag}}+E_{\textrm{ther}}$)/$|E_{\textrm{grav}}|$), of which positive (or negative) value indicates that the core is gravitationally unbound (or bound).

The maximum mass of dense cores increases from $t$ = 1.3 Myr to the shock-crossed epoch of Medium cloud, $t$ = 1.8 Myr. 
Before $t$ = 1.8 Myr, no bound core with mass more than 10 $M_{\odot}$ (hereafter, we call such cores the massive bound cores) is formed.

After the shock-crossed epoch of Medium cloud, the total number of dense cores and dense cores more than 10 $M_{\odot}$ (hereafter, we call such cores the massive cores) decreases with time. 
The total number of dense cores decreases from twenty-one at $t$ = 1.8 Myr to twelve at $t$ = 2.5 Myr, and the number of massive cores decreases from four at $t$ = 1.8 Myr to one at $t$ = 2.5 Myr.
These decreases appear during the gas expansion of the leading part of the shocked region which occurs after the shock-crossed epoch of Medium cloud, as shown in bottom panels in Figure \ref{fig:dens_slice_M20strong}. 
Since the total pressure of the leading part of the shocked region decreases during this gas expansion, highly unbound massive cores lose their mass as shown later in this subsection.

At $t$ = 3.1 Myr, there are two massive bound cores, as shown in Figure \ref{fig:cmf_M20strong}. We expect formation of a very small number of massive stars in M20. We note that total mass of dense cores agrees well with $M_{\mathrm{high,tot}}$ shown in Figure \ref{fig:highdenseCCC}, since we use same $\rho_{\textrm{th}}$ for the selection of dense cores and the high-density gas regions. 
We show the CMDs for M10 at $t$ = 1.3, 1.8, 2.1, and 3.1 Myr in Figure \ref{fig:cmf_M10strong} for comparison. 
One dense core is formed at $t$ = 1.3 Myr, while many dense cores are already formed in M20 at the same epoch, as shown in Figure \ref{fig:cmf_M20strong}. 
This is because the shocked region accumulates gas faster in M20 than in M10 due to the higher collision speed in M20 than in M10. 

\begin{figure*}
    \begin{center}
    \includegraphics[width=\textwidth]{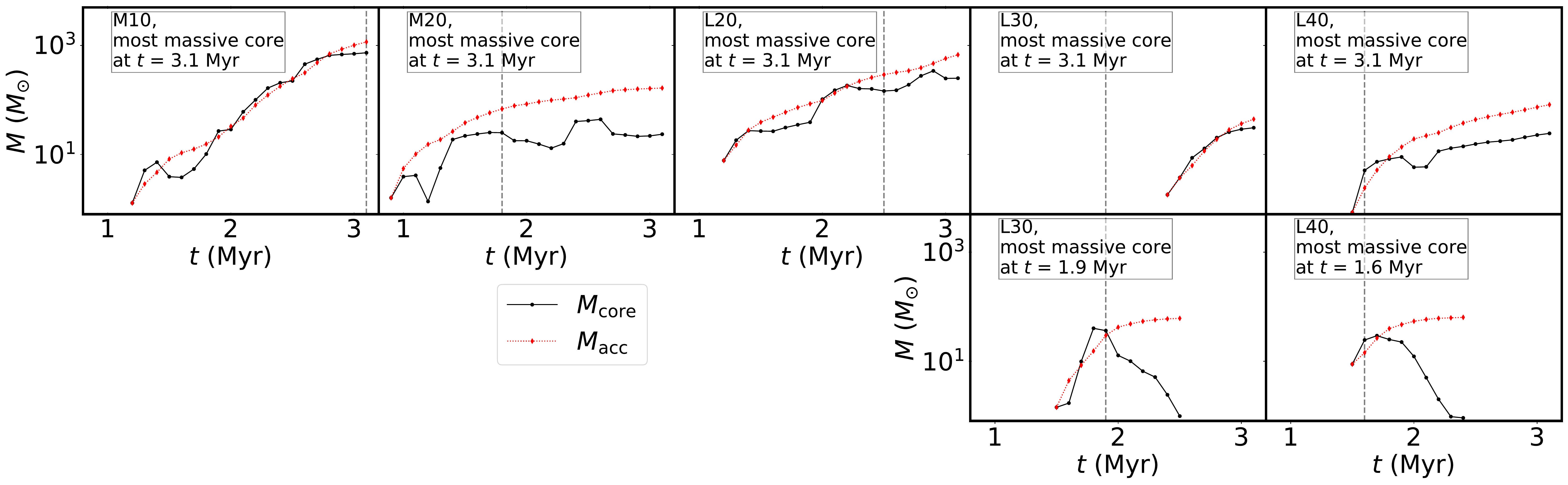}
    \end{center}
\caption{Top panels: Mass evolution of the core which is most massive at $t$ = 3.1 Myr in all CCC models shown by solid black lines. Bottom panels: Mass evolution of most massive core at the shock-crossed epoch of Large cloud in L30 and L40 shown by solid black lines. The dotted red lines show $M_{\textrm{acc}}$. The details of this mass estimate are given in the main text. The vertical dashed line shows the shock-crossed epoch of Medium cloud in M10 and M20 and the shock-crossed epoch of Large cloud in L20, L30, and L40.}
\label{fig:accretionM20} 
\end{figure*}

At the shock-crossed epoch of Small cloud ($t$ = 1.3 Myr for M20 as shown in Figure \ref{fig:cmf_M20strong}, and $t$ = 2.1 Myr for M10 as shown in Figure \ref{fig:cmf_M10strong}), the number of massive cores is lower in M20 than M10, although the shocked regions contain roughly the same mass. 
This may be due to
difference of mass evolution of dense cores in both models. 

At the shock-crossed epoch of Medium cloud ($t$ = 1.8 Myr for M20 as shown in Figure \ref{fig:cmf_M20strong}, and $t$ = 3.1 Myr for M10 as shown in Figure \ref{fig:cmf_M10strong}), no massive bound core is formed in M20, whereas four massive bound cores are formed in M10. The maximum mass of dense core is significantly lower in M20 than M10 at this stage.
The total mass of dense cores agrees well with $M_{\mathrm{high,tot}}$ in both models.

We examine evolution of dense cores in both models. Figure \ref{fig:energies2} shows time evolution of masses and gravitational boundness ratios of ten most massive dense cores at the shock-crossed epoch of Medium cloud in M20 (left panels) and M10 (right panels). We use the same color for lines to show evolution of the cores which finally merge into one core. This figure clearly shows that mass evolution of the cores is very different between M20 and M10.
In M20, mass evolution of the cores is suppressed after the shock-crossed epoch of Medium cloud, $t$ = 1.8 Myr, and they are gravitationally unbound at $t$ = 3.1 Myr, except for cores shown by red lines and green line.
In M10, the cores roughly increase their mass with time to $t$ = 3.1 Myr, and five cores among them are gravitationally bound at this epoch.

We study the reason for the destruction of cores in M20 by comparing mean magnetic pressure of the shocked region with non-thermal pressure at the surfaces of dense cores estimated by 
\begin{equation}
P_{\textrm{sur}} = ((E_{\textrm{turb}}+E_{\textrm{mag}})\rho_{\textrm{th}})/M_{\textrm{core}}
\end{equation}

The bottom panels of Figure \ref{fig:energies2} show $P_{\textrm{sur}}$ of the cores.
The dashed orange line shows the mean magnetic pressure of the shocked region, $P_{\textrm{mag,sh}}$ = ${B_{\textrm{sh}}}^2/8\pi$. 
The dashed and dotted grey lines show typical ram pressures to the shocked region by the cloud medium, $P_{\textrm{ram,C}}$ = $\rho_0{v_{\textrm{sh}}}^2$, and by the ambient medium, $P_{\textrm{ram,A}}$ = $\rho_{\textrm{amb}}{v_{\textrm{sh}}}^2$, respectively, where $\rho_{\textrm{amb}}$ is the density of the ambient medium and $v_{\textrm{sh}}$ is mean speed of the shocked region. 
For these estimations, we select the shocked region as a group of cells with density $>$ 5$\rho_0$ and with a positive $x$-component of velocity. $P_{\textrm{mag,sh}}$ is a volume-weighted average quantity, and $v_{\textrm{sh}}$ is a mass-weighted average quantity of the $x$-component of velocity.
 
In M20, $P_{\textrm{mag,sh}}$ is comparable to $P_{\textrm{ram,C}}$ before the shock-crossed epoch of Medium cloud. This means that $P_{\textrm{mag,sh}}$ is the dominant supporting presure of the shocked region. $P_{\textrm{mag,sh}}$ decreases faster than $P_{\textrm{ram,C}}$ after the shock-crossed epoch of Medium cloud.
After this epoch, the ram pressure which acts on the leading part of the shocked region changes from $P_{\textrm{ram,C}}$ to $P_{\textrm{ram,A}}$ which is much smaller than $P_{\textrm{ram,C}}$.
This decrease of the ram pressure induces the expansion of the leading part and decrease of $P_{\textrm{mag,sh}}$. The cores which are highly unbound and have $P_{\textrm{sur}}$ higher than $P_{\textrm{mag,sh}}$ at the shock-crossed epoch of Medium cloud are mostly destroyed. We show this in the next paragraph by comparing the time evolution of five most massive cores at $t$ = 1.8 Myr in M20. In M10, the cores are not destroyed, since the change of $P_{\textrm{mag,sh}}$ is small.

We compare time evolution of five most massive cores at $t$ = 1.8 Myr in M20 (green, red, blue, red, and pink lines in descending order of core mass shown in Figure \ref{fig:energies2}). The most massive (green line), third most massive (blue line), and fifth most massive (pink line) cores are highly unbound at $t$ = 1.8 Myr as the gravitational boundness ratio $\sim$ 3.5. 
These three cores significantly lose their mass from $t$ = 1.8 Myr to 2.0 Myr. 
The second and fourth most massive cores (both shown by red lines) do not change their mass considerably in this period. These two cores are not highly unbound at $t$ = 1.8 Myr as their gravitational boundness ratio $\sim$ 1.9 and 1.2.

The top, left two panels in Figure \ref{fig:accretionM20} show the mass evolution of the most massive bound core, $M_{\textrm{core}}$, at $t$ = 3.1 Myr and its estimated accreted mass, $M_{\textrm{acc}}$ (see equation (\ref{eq:accretedmass}) in Section \ref{model-densecore} for details) in M20 and M10. 
We show their evolution from $t$ = 0.9 Myr for M20 and from $t$ = 1.2 Myr for M10.
The accreted mass, $M_{\textrm{acc}}$, well reproduces $M_{\textrm{core}}$ before the shock-crossed epochs of Medium cloud, $t$ = 1.8 Myr in M20 and $t$ = 3.1 Myr in M10. 
After the shock-crossed epoch in M20, $M_{\textrm{core}}$ becomes smaller than $M_{\textrm{acc}}$. This indicates that the core losses its mass by the expansion to surrounding region of the core. Such evolution of gravitationally unbound cores explains the suppression of total mass evolution of high-density gas regions in M20.

We find that highly unbound cores easily lose their mass, and gas accretion to such dense cores is suppressed during expansion of the shocked region. We expect that a target cloud of larger size which is expected to have a later shock-crossed epoch of target cloud than M20 will have more time for a mass increase of dense cores by accretion. Such a target cloud will favor formation of massive bound cores for 20 km $^{-1}$ collision speed. For the above reason, we simulate the collision between Small and Large clouds, since the shock-crossed epoch of Large cloud is later than that of Medium cloud for the same collision speed. We give the numerical results of collision of Small and Large clouds in Section \ref{cld_size_effect}.

\subsection{Collision of Small and Large clouds}\label{cld_size_effect}
In this section, we show simulation results of Small and Large, magnetized, colliding clouds. 
In Section \ref{section:L20}, we show simulation results of 20 km s$^{-1}$ collision speed, L20, and compare it with M20. 
In Section \ref{section:L_densecore}, we compare formation and evolution of dense cores in both models. In Section \ref{L30strong}, we show results of 30 km s$^{-1}$ (L30) and 40 km s$^{-1}$ (L40).

\subsubsection{Time evolution of L20 and its comparison with M20}\label{section:L20}
\begin{figure*}
    \begin{center}
    \includegraphics[width=0.9\textwidth]{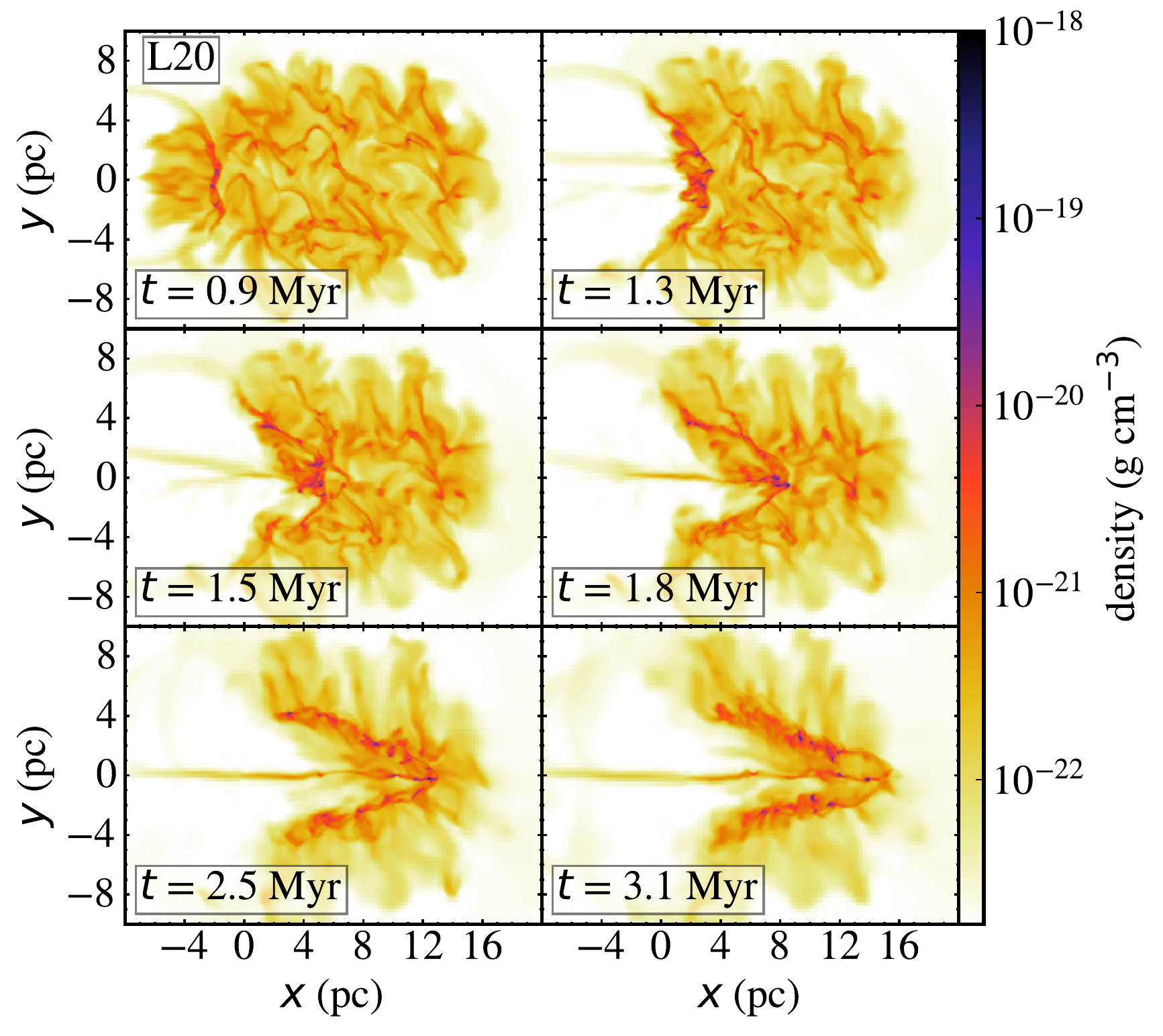}
    \end{center}
    \caption{Slice plots of the gas density in $z$ = 0 pc at $t$ = 0.9, 1.3, 1.5, and 1.8, 2.5, and 3.1 Myr in L20. The color bar shows gas density.}
    \label{fig:dens_slice_L20strong} 
\end{figure*}

\begin{figure*}
    \begin{center}
    \includegraphics[width=1\textwidth]{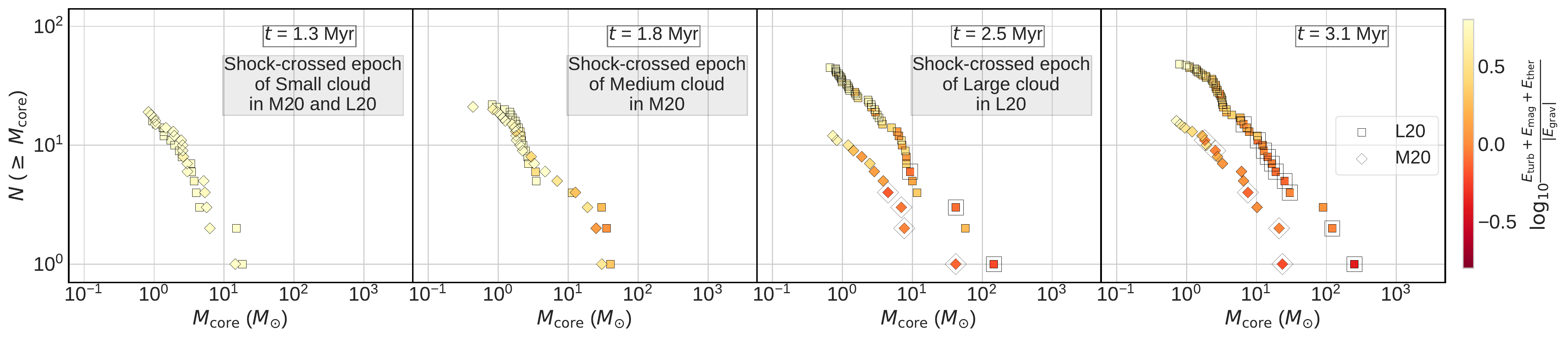}
    \end{center}
    \caption{Cumulative dense core mass distributions shown by filled squares and diamonds at $t$ = 1.3, 1.8, 2.5, and 3.1 Myr for L20 and M20, respectively. The gravitationally bound cores are marked by larger, open squares and diamonds for L20 and M20, respectively.}
    \label{fig:cmf_L20strong} 
\end{figure*}

Figure \ref{fig:dens_slice_L20strong} shows time evolution of the gas density structures of L20.
At $t$ = 0.9, 1.3, 1.5, and 1.8 Myr, the shocked regions in L20 have similar structures to those at same epochs in M20 (see Figure \ref{fig:dens_slice_M20strong}). We note that same initial turbulent velocities are used to generate turbulent structures. 
$M_{\mathrm{high,tot}}$ in L20 is similar to M20 before $t$ = 1.8 Myr, as shown in left panel of Figure \ref{fig:highdenseCCC}. 

After $t$ = 1.8 Myr, $M_{\mathrm{high,tot}}$ monotonically increases with time in L20, while it does not increase with time in M20, as shown in Figure  \ref{fig:highdenseCCC}. 
This difference is due to different evolution of the shocked regions after $t$ = 1.8 Myr. 
From $t$ = 1.8 to 2.5 Myr, the shocked region moves in Large cloud in L20, as shown in Figure \ref{fig:dens_slice_L20strong}. The mass of the shocked region increases by sweeping up the gas in Large cloud. In the same period, the shocked region expands, and mass evolution of cores is suppressed, since a leading part of the shocked region has already moved in the ambient medium in M20, as shown in Figure \ref{fig:dens_slice_M20strong}. 

At $t$ = 2.5 Myr, several high-density regions are formed in the shocked region in L20.
After $t$ = 2.5 Myr, a leading part of the shocked region goes through the right edge of Large cloud.
This leads to expansion of the shocked region near Large cloud's right edge. 
Figure \ref{fig:dens_slice_L20strong} shows expansion of the leading part of the shocked region from $t$ = 2.5 Myr to $t$ = 3.1 Myr.
In this period, $M_{\mathrm{high,tot}}$ still increases with time, as shown in Figure \ref{fig:highdenseCCC}.

\subsubsection{Dense core formation and evolution}\label{section:L_densecore}

Figure \ref{fig:cmf_L20strong} shows CMDs at $t$ = 1.3, 1.8, 2.5, and 3.1 Myr by filled squares for L20 and diamonds for M20 for comparison. The CMDs in both models are very similar to each other at $t$ = 1.3 and 1.8 Myr. This result is reasonable, since both the models have similar $M_{\mathrm{high,tot}}$ till $t$ = 1.8 Myr, as described in Section \ref{section:L20}. 

After $t$ = 1.8 Myr, difference of the CMDs between L20 and M20 becomes large. 
At $t$ = 2.5 Myr, two massive bound cores are formed in L20, whereas there is only one such massive bound core in M20. 
At $t$ = 3.1 Myr, nine massive bound cores are formed in L20, whereas there are only two massive bound cores in M20. The total number of dense cores in L20 is larger than that in M20 after $t$ = 1.8 Myr.

These differences in the total number of massive bound cores and the total number of dense cores between L20 and M20 can be explained by the different evolution of the dense cores in the shocked regions. 
The longer time available for the dense cores to accumulate gas in the shocked region favors massive bound core formation in L20 than M20. 

We explain this using two examples of most massive core at $t$ = 3.1 Myr in L20 and M20 as follows.
The time evolution of mass of the core, $M_{\textrm{core}}$, which is most massive at $t$ = 3.1 Myr can be explained by the estimated accreted mass, $M_{\textrm{acc}}$, up to the shock-crossed epoch of Large cloud, $t$ = 2.5 Myr, in L20 and up to the shock-crossed epoch of Medium cloud, $t$ = 1.8 Myr, in M20, as shown in top panels in Figure \ref{fig:accretionM20}. 
Since the shock-crossed epoch of Large cloud in L20 occurs after the shock-crossed epoch of Medium cloud in M20, dense cores get higher mass by the accretion for a longer time in L20
than that in M20.

We have shown that many massive bound cores form in L20.  For the same cloud models, we extend our simulations to higher collision speeds, 30 and 40 km s$^{-1}$, to study a collision speed limit for massive bound core formation. In next Section \ref{L30strong}, we show results of collisions of Small and Large clouds with these speeds (L30 and L40).

\subsubsection{L30 and L40} \label{L30strong} 
\begin{figure*}
    \begin{center}
    \includegraphics[width=0.85\textwidth]{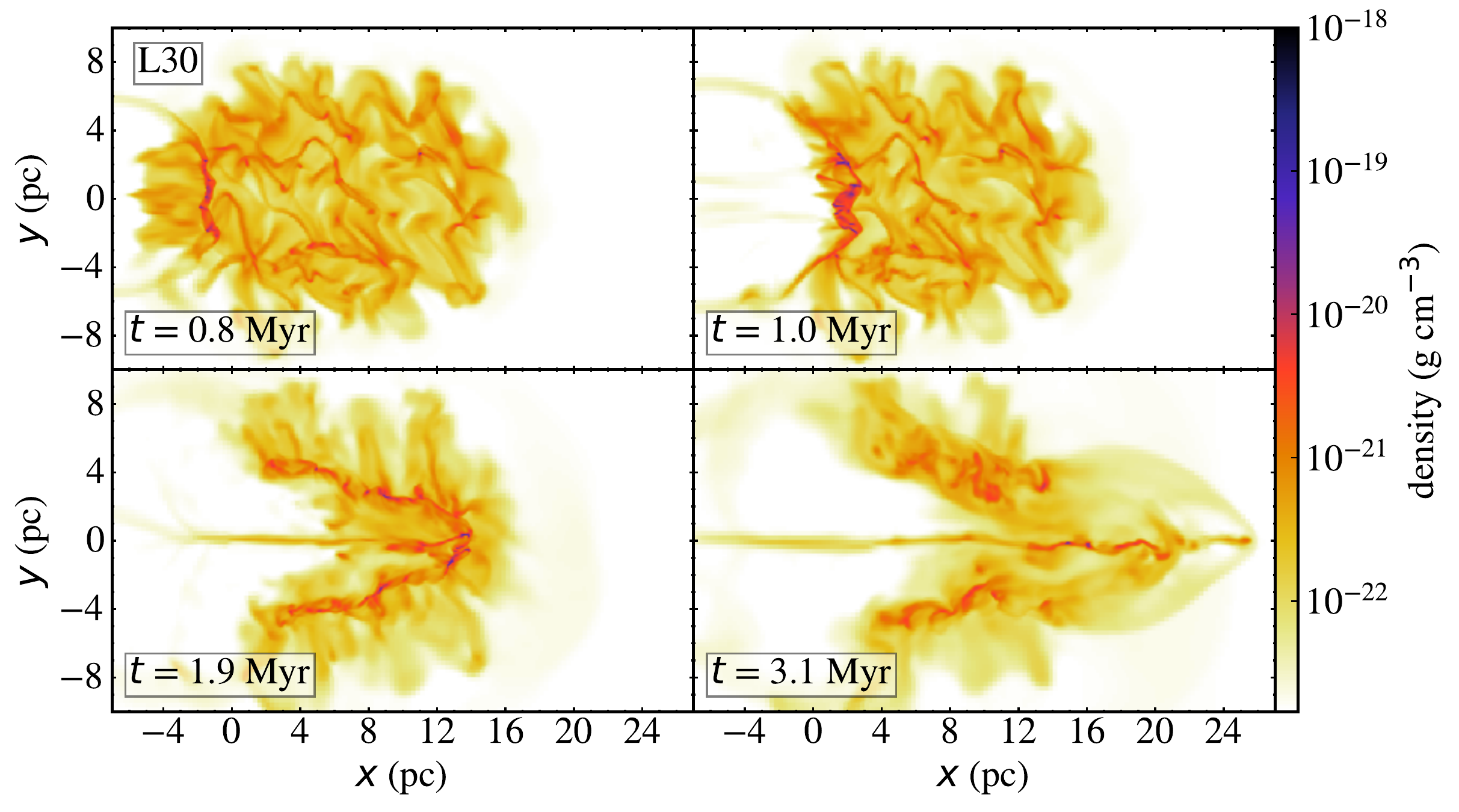}
    \end{center}
    \caption{Slice plots of the gas density in $z$ = 0 pc at $t$ = 0.8, 1.0, 1.9, and 3.1 Myr in L30. The color bar shows gas density.}
    \label{fig:dens_slice_L30strong}
\end{figure*}

\begin{figure*}
    \begin{center}
    \includegraphics[width=1\textwidth]{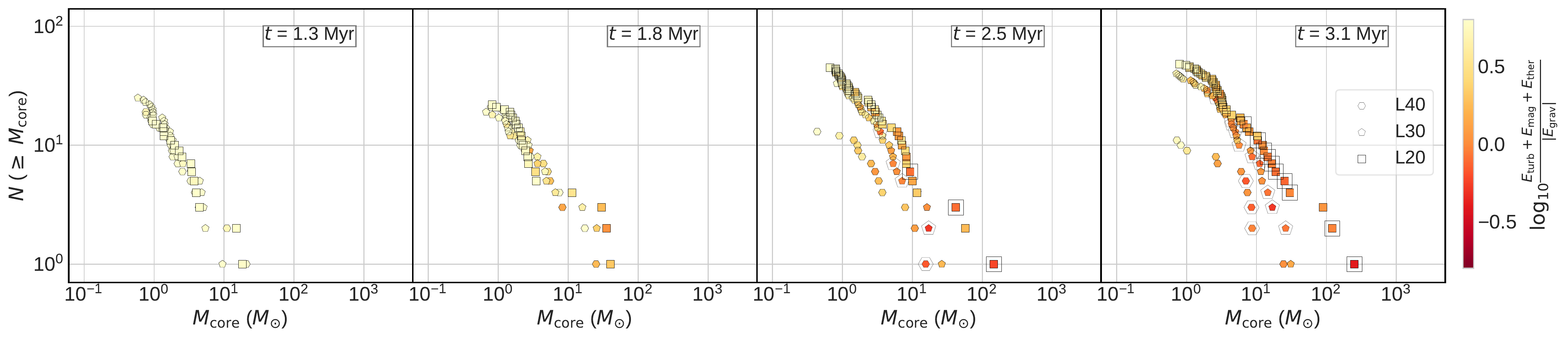}
    \end{center}
    \caption{Same as in Figure \ref{fig:cmf_L20strong}, but for L40, L30, and L20.}
    \label{fig:cmf_L30strong}
\end{figure*}

\begin{figure*}
    \begin{center}
    \includegraphics[width=1\textwidth]{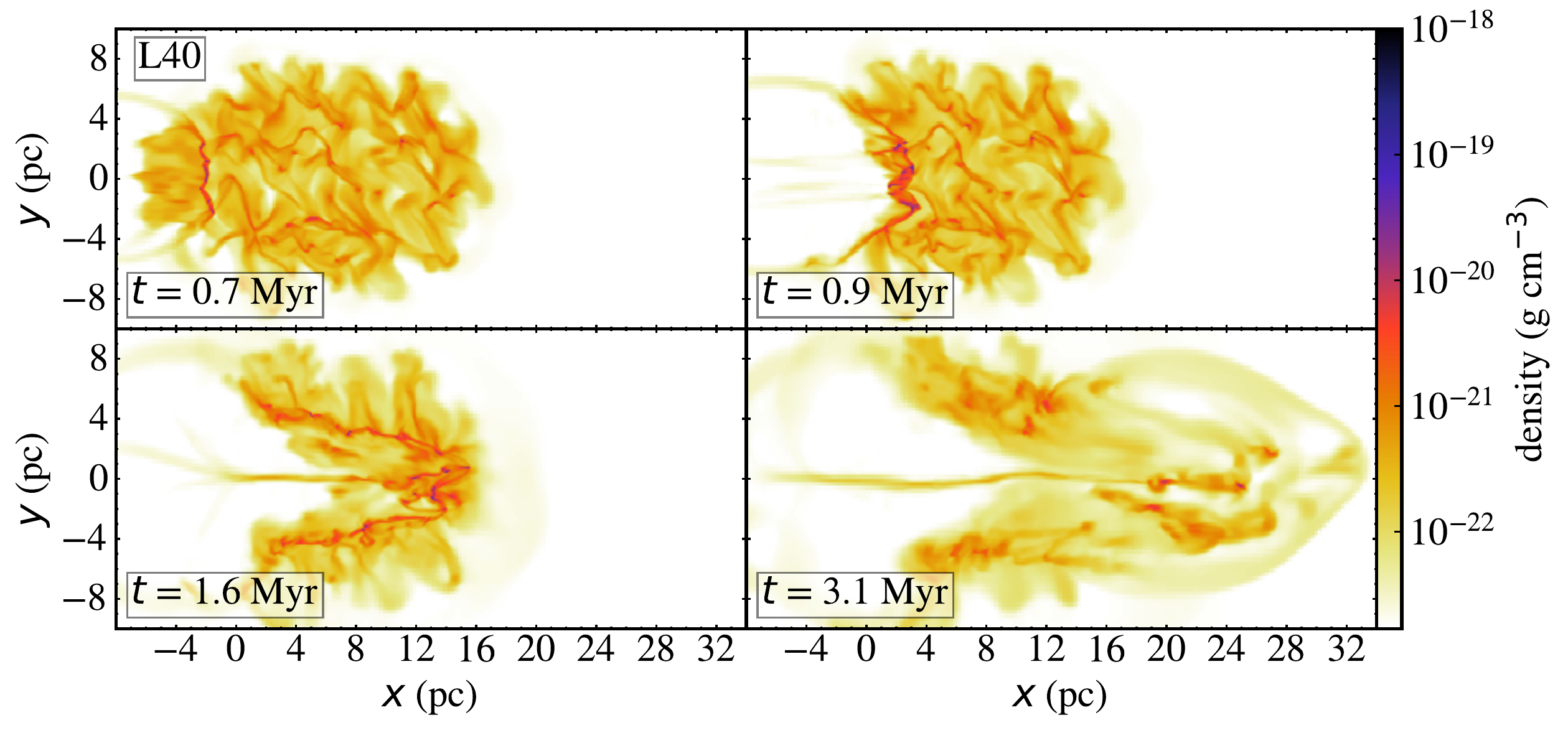}
    \end{center}
    \caption{Slice plots of the gas density in $z$ = 0 pc at $t$ = 0.7, 0.9, 1.6, and 3.1 Myr in L40. The color bar shows gas density.}
    \label{fig:dens_slice_L40strong}
\end{figure*}
Figure \ref{fig:dens_slice_L30strong} shows time evolution of the gas density structures of L30.
The panels in this figure show early stage of the collision between Small and Large clouds at $t$ = 0.8 Myr, the shocked region at the shock-crossed epoch of Small cloud at $t$ = 1.0 Myr, the shocked region at the shock-crossed epoch of Large cloud at $t$ = 1.9 Myr, and the end of simulation at $t$ = 3.1 Myr.

In the early stage of the collision (top left panel in Figure \ref{fig:dens_slice_L30strong}) and in the stage of the shock-crossed epoch of Small cloud (top right panel in Figure \ref{fig:dens_slice_L30strong}), the structure of the shocked region is similar to that in the corresponding stages of L20. 
In those stages, $M_{\mathrm{high,tot}}$ in L30 are similar to those in L20, as shown in middle panel of Figure \ref{fig:highdenseCCC}.

In the stage of the shock-crossed epoch of Large cloud (bottom left panel in Figure \ref{fig:dens_slice_L30strong}), the high-density gas regions in the shocked region are concentrated to the collision axis. 
$M_{\mathrm{high,tot}}$ in this stage is lower in L30 than that in L20, as shown in Figure \ref{fig:highdenseCCC}. 
This difference shows less accumulation of dense gas in the high-density gas regions in L30 than L20. This occurs because of less time in L30 than L20 for the high-density gas regions to accumulate gas by the flow along the deformed magnetic fields.

After the shock-crossed epoch of Large cloud, a leading part of the shocked region moves in the ambient medium and expands.
$M_{\mathrm{high,tot}}$ hardly increases in L30, as shown in Figure \ref{fig:highdenseCCC}. 
At $t$ = 3.1 Myr, the leading part of the shocked region has significantly expanded in the ambient medium, as shown in Figure \ref{fig:dens_slice_L30strong}. During this expansion, the core which is most massive at the shock-crossed epoch of Large cloud is destroyed, as shown in the bottom left panel in Figure \ref{fig:accretionM20}. 
The core which is most massive at $t$ = 3.1 Myr grows in the non-expanding part of the shocked region. 
The $M_{\textrm{acc}}$ well reproduces $M_{\textrm{core}}$ of this core, as shown in the top panel in Figure \ref{fig:accretionM20}.
This core is not gravitationally bound at $t$ = 3.1 Myr. 
$M_{\mathrm{high,tot}}$ in L30 is much smaller at this epoch than L20, as shown in Figure \ref{fig:highdenseCCC}. 

Figure \ref{fig:cmf_L30strong} shows CMDs at $t$ = 1.3, 1.8, 2.5, and 3.1 Myr in L30 and L20. The figure shows that CMDs are not so different between these models at $t$ = 1.3 and 1.8 Myr. At $t$ = 2.5 and 3.1 Myr, the number of massive bound cores in L30 is smaller than that in L20.
The massive bound cores formed in L30 are significantly less massive than those in L20.
These massive bound cores in L30 are not in the expanding part of the shocked region, and they grow by the accretion.

Figure \ref{fig:dens_slice_L40strong} shows time evolution of the gas density structure of L40.
In early phase in L40, more high-density regions are formed than in L30, as shown by $M_{\mathrm{high,tot}}$ in middle panel of Figure \ref{fig:highdenseCCC}. 
This is due to higher collision speed in L40 than L30.
At the shock-crossed epoch of Large cloud, $M_{\mathrm{high,tot}}$ in L40 is similar to that in L30, as shown in Figure \ref{fig:highdenseCCC}. 
In the stage near the shock-crossed epoch of Large cloud (bottom left panels in Figure \ref{fig:dens_slice_L40strong} and Figure \ref{fig:dens_slice_L30strong}), the shocked region is more dispersed in L40 than L30. 
This may be due to large irregular motions in the shocked region produced by NTSI by higher collision speed in L40 than L30.

After the shock-crossed epoch of Large cloud, a leading part of the shocked region moves in the ambient medium and expands. 
Then, $M_{\mathrm{high,tot}}$ decreases, saturates later, and is significantly lower than that in L20 and L30 at $t$ = 3.1 Myr, as shown in Figure \ref{fig:highdenseCCC}. The core which is most massive at the shock-crossed epoch of Large cloud decreases its mass with time after that epoch, as shown in the bottom right panel in Figure \ref{fig:accretionM20}.
At $t$ = 3.1 Myr, there is one massive core formed in the non-expanding part of the shocked region, as shown in the top right panel in Figure \ref{fig:accretionM20}. 
This core is not gravitationally bound at $t$ = 3.1 Myr. 

The CMDs show that the number of dense cores and maximum mass of gravitationally bound dense cores are smaller in L40 than L20 and L30 at $t$ = 3.1 Myr, as shown in Figure \ref{fig:cmf_L30strong}. 
These differences become large after the shock-crossed epoch of Large cloud in L40, $t$ = 1.6 Myr. This significantly small number of dense cores in L40 than L30 and L20 may be due to large irregular motion of dense cores. Further discussion on this is given in Section \ref{discuss_1_cols}.
\section{Discussion}\label{discuss}
\subsection{Collision speed and target cloud size effects}\label{discuss_1_cols}

\begin{figure*}
    \begin{center}
    \includegraphics[width=1\textwidth]{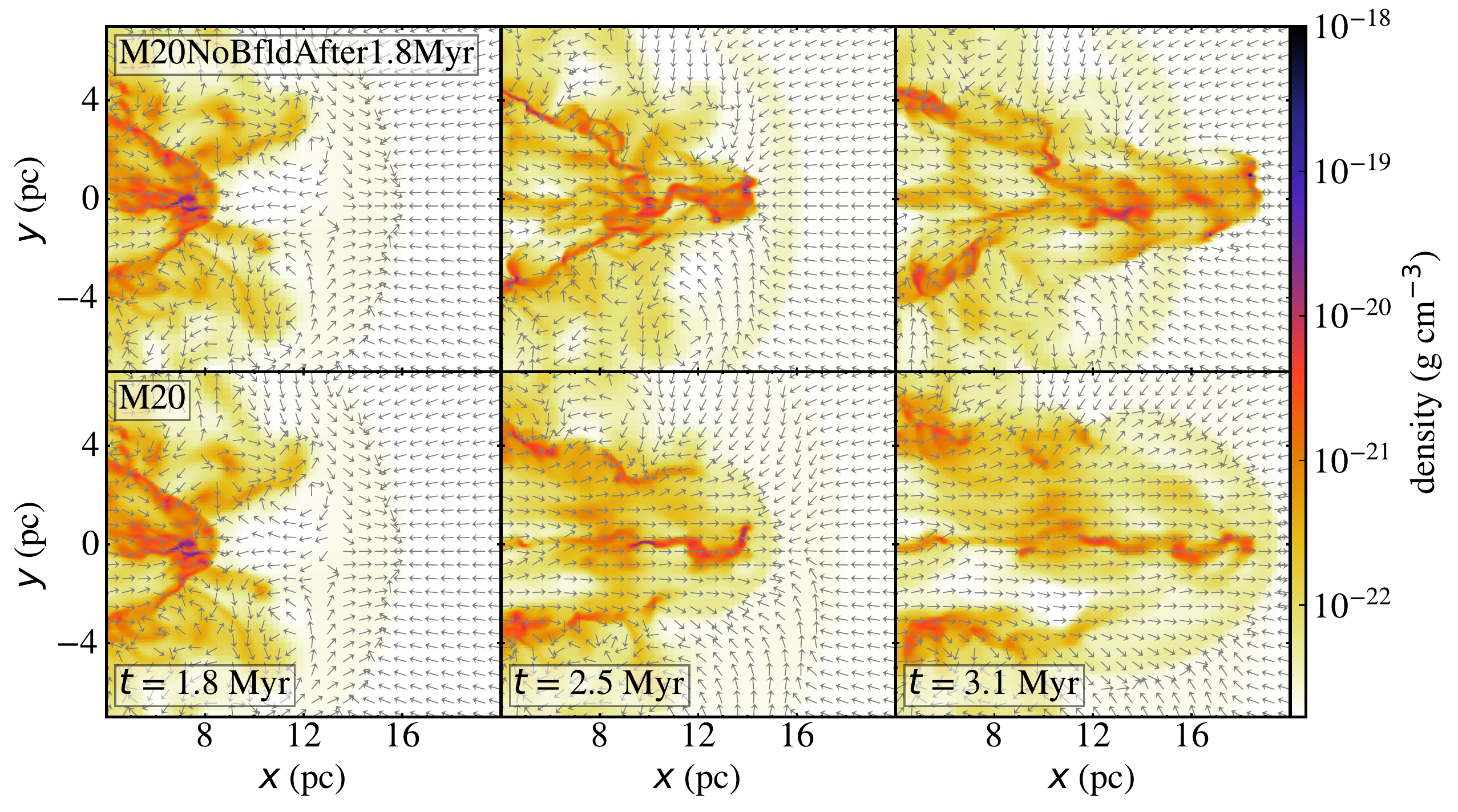}
    \end{center}
    \caption{Slice plots of the gas density in $z$ = 0 pc at $t$ = 1.8, 2.5, and 3.1 Myr in a modified model of M20 which is restarted from $t$ = 1.8 Myr without magnetic field (top panels) and in model M20 (bottom panels). The arrows show unit vectors of $(v_x, v_y)/\sqrt{v_x^2+v_y^2}$, where $v_x$, $v_y$, and $v_z$ are $x$-, $y$-, and $z$-components of velocity in the frame of simulation box, respectively. The color bar shows gas density.}
    \label{fig:dens_slice_extramodel}
\end{figure*}

\begin{figure*}
    \begin{center}
    \includegraphics[width=1\textwidth]{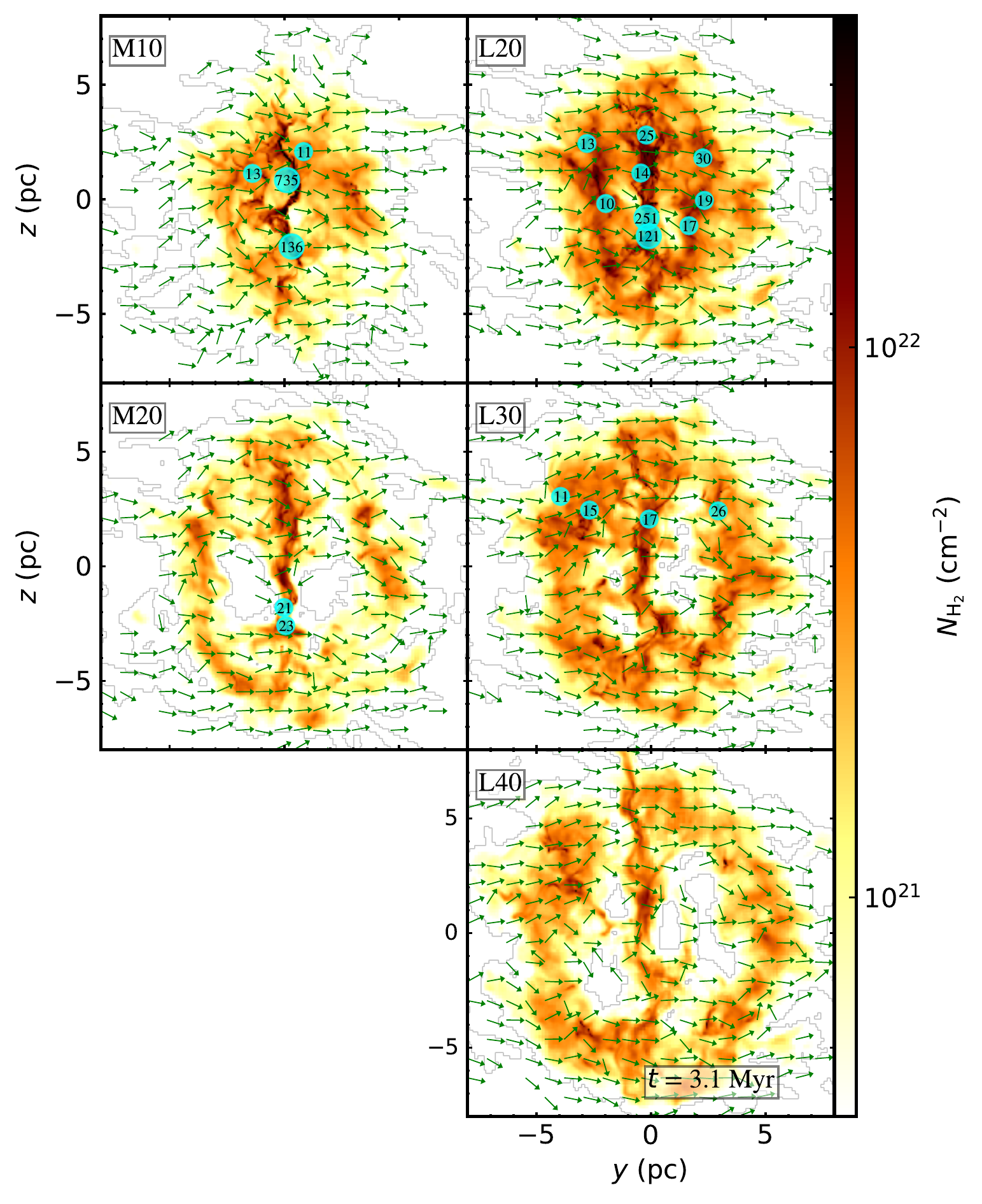}
    \end{center}
    \caption{H$_2$ column density, $N_{\mathrm{H}_2}$, along the collision axis in the $y$-$z$ plane at $t$ = 3.1 Myr for all CCC models. Markers show positions of massive bound cores and a number on each marker shows its mass in unit of $M_{\odot}$. Large marker is used  for massive bound cores greater than 100 $M_{\odot}$. The color bar shows the column density and the arrows show unit vectors along the mass-weighted magnetic fields averaged along the collision axis.}
    \label{fig:column_CCC} 
\end{figure*}

\begin{figure*}
\begin{center}
\includegraphics[width=0.8\textwidth]{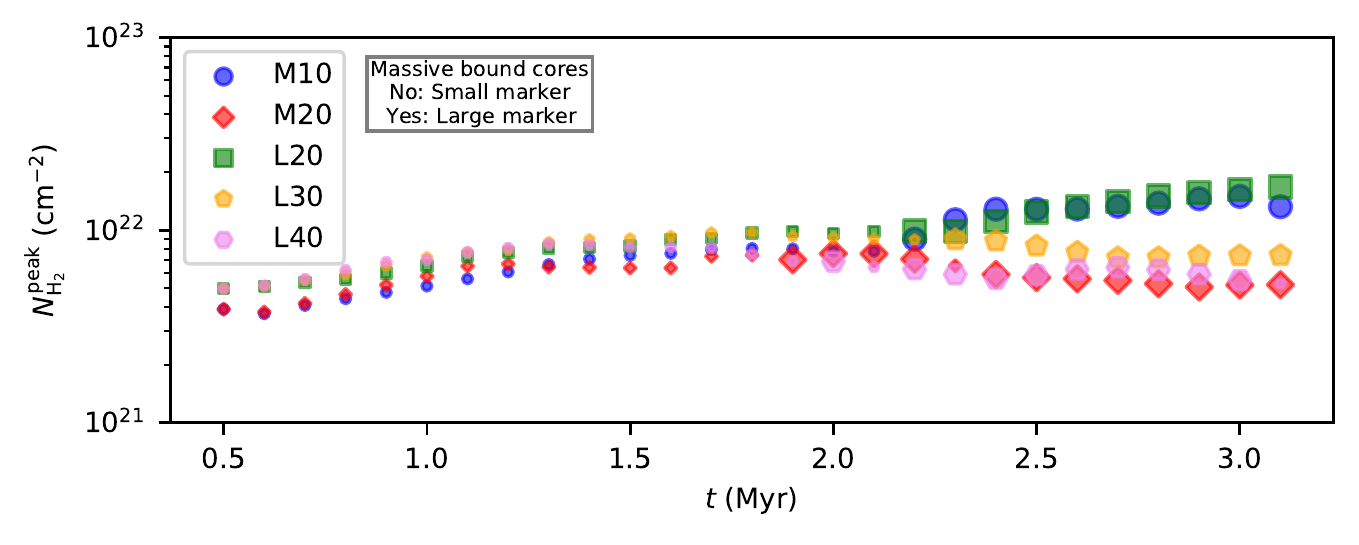}
\includegraphics[width=0.85\textwidth]{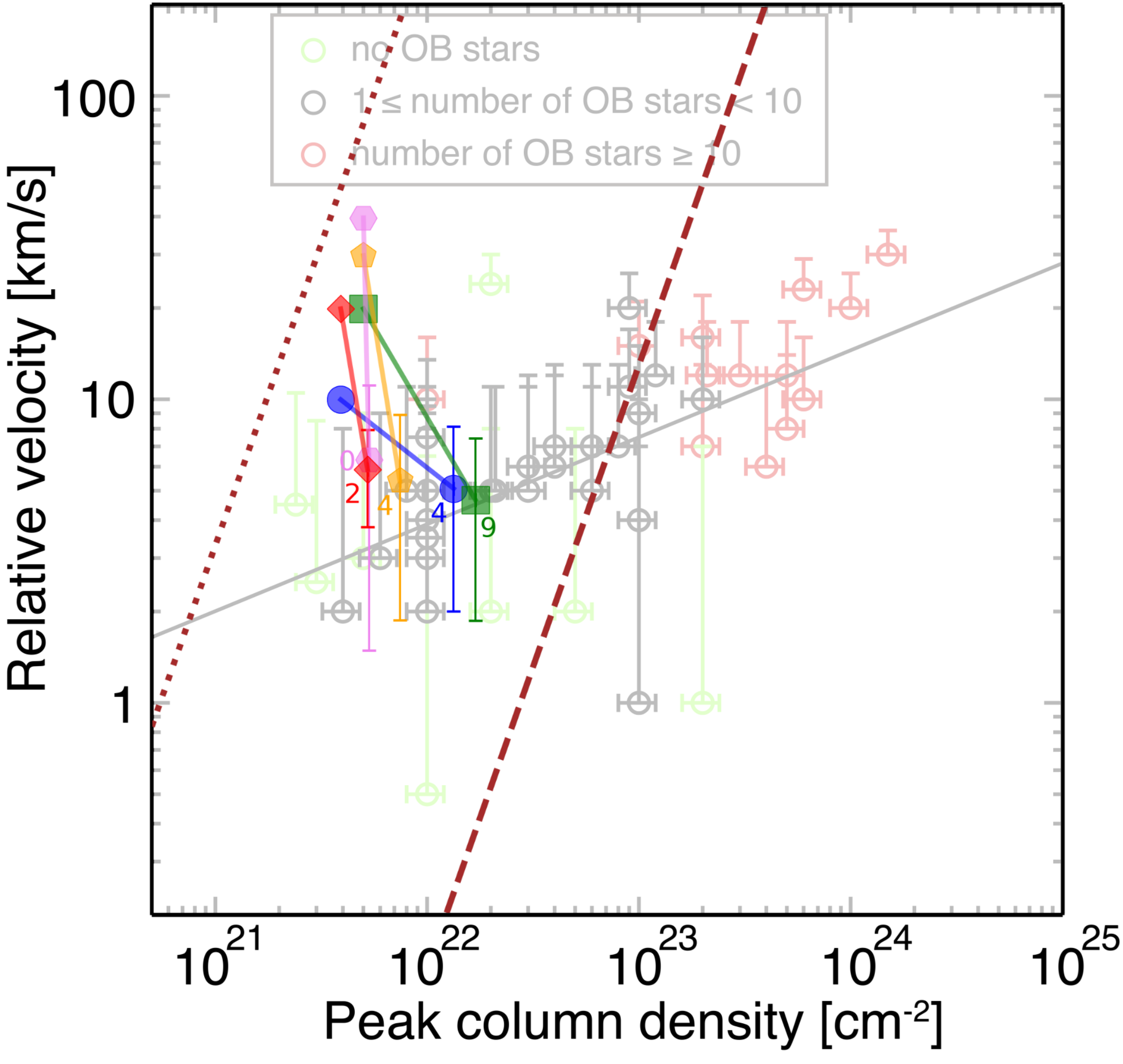}
\end{center}
\caption{Top panel: Time evolution of peak H$_2$ column density, $N_{\mathrm{H}_2}^{\mathrm{peak}}$, measured along the collision axis with 1 pc spatial resolution from $t$ = 0.5 to 3.1 Myr for CCC models. Larger markers are used for epochs after the massive bound core formation. 
Bottom panel: $N_{\mathrm{H}_2}^{\mathrm{peak}}$ and the collision speed at $t$ = 0.5 Myr and the estimated collision speed with error bars at final simulation epoch ($t$ = 3.1 Myr) of CCC models with same markers used in the top panel on the translucent image of Figure 9(a) in \citet{2021PASJ...73S..75E}. The top and bottom of the same markers show $N_{\mathrm{H}_2}^{\mathrm{peak}}$ at $t$ = 0.5 and 3.1 Myr, respectively. The total number of massive bound cores at $t$ = 3.1 Myr is written beside the bottom  markers. Analytical conditions given by equation (\ref{eq:dis7}) are shown by dotted and dashed lines. The method to estimate collision speed at $t$ = 3.1 Myr is given in main text.}
\label{fig:discuss} 
\end{figure*}

We have shown that formation of massive (> 10 $M_{\odot}$) bound cores is well suppressed in the collision speed of 20 km s$^{-1}$ than 10 km s$^{-1}$ for collision between Small and Medium clouds. 
The number of dense cores decreases with time in M20 after the shock-crossed epoch of Medium cloud.
For the collision speed of 20 km s$^{-1}$, if we increase the target cloud size from that of Medium cloud to Large cloud, the collision between Small and Large clouds results in a larger number of dense cores than that in the collision between Small and Medium clouds.
We discuss the reason why the target cloud size affects the formation of dense cores.

\citet{2014ApJ...792...63T} and \citet{2018PASJ...70S..58T} using hydrodynamic simulations demonstrated that the higher collision speed suppresses the mass increase of massive part of CMD in their colliding clouds with no magnetic field. 
The suppression of the mass increase is due to a shorter accretion time scale in a higher speed collision for the same colliding clouds.
The accretion to dense cores continues, if dense cores remain in the shock region produced by CCC. More massive cores are harder to decelerate and go through the shocked region earlier than less massive cores. 
In this way, mass accretion stops earlier for more massive cores than less massive cores.
This process produces a CMD with a steeper power index of $M_{\textrm{core}}$ in the higher mass part of CMD.
We expect that a similar process occurs in M20 and that a higher collision speed of more than 20 km s$^{-1}$ results in suppression of massive bound core formation in the collision of Small and Large clouds.

To study collision speed limit for massive bound core formation in collision of Small and Large clouds, we have simulated their collision with speeds of 30 km s$^{-1}$ (L30) and 40 km s$^{-1}$ (L40). The number of massive bound cores decreases with increasing collision speed, and there are no massive bound cores in L40 at the end of the simulation. The change in the CMD mainly appears after the shock-crossed epoch of Large cloud.
After the shock-crossed epoch, the leading part of the shocked region moves in the ambient medium, and this part rapidly expands with time due to the small ram pressure by the ambient medium. 
Total pressure in the leading part decreases due to this expansion.
Highly unbound cores in this part disintegrate due to this rapid decrease of the total pressure.
Since \citet{2018PASJ...70S..58T} did not show such expansion of the shocked region, this expansion is mainly caused by the magnetic pressure in the shocked region. We confirm this by restarting M20 at the shock-crossed epoch of Medium cloud ($t$ = 1.8 Myr) without magnetic field. Top panels in Figure \ref{fig:dens_slice_extramodel} show that this modified model has no expansion of the shocked region at later epochs, $t$ = 2.5 and 3.1 Myr.

Another possible process in L40 to suppress the mass increase of dense cores is large irregular motions of dense cores in the shocked region.
The speeds of irregular motions of dense cores are as high as 10.4, 8.0, and 4.6 km s$^{-1}$ at the shocked-crossed epoch of Small cloud in L40, L30, and L20, respectively. It is possible that such large irregular motions in L40 can disturb the mass growth of dense cores.
Such large irregular motions can be due to NTSI.
In Paper I, we estimate a magnetic field that can suppress NTSI for a given collision speed.   
The collision speed in L40 is high enough to overcome the suppression effect of magnetic field of 4 \textmu G on NTSI, as discussed in Paper I.

CCCs with high speeds are suggested in bar regions of barred galaxies. \citet{2014MNRAS.445L..65F} and \citet{2020MNRAS.494.2131F} have shown via galaxy-scale simulations of a barred galaxy that a large fraction of colliding clouds in the bar region has collision speeds of more than 30 km s$^{-1}$. 
They pointed out that CCCs with such high collision speeds can explain the low star formation efficiency observed in the bar regions in some barred galaxies \citep[e.g.][]{2010ApJ...721..383M,2021MNRAS.502.2238M}, if massive star formation is suppressed in CCCs with high collision speeds due to suppression of massive bound core formation, which is demonstrated by hydrodynamic simulations in \citet{2014ApJ...792...63T} and \citet{2018PASJ...70S..58T} and also by MHD simulations in this paper.

We apply the least square fits using equation (\ref{cmf_fit}) to the CMDs of massive cores at $t$ = 3.1 Myr for all CCC models except L40 which has only one massive core. 
The power index of core mass function, $\gamma$, is $\sim$ 1.3 in M10, $\sim$ 2.0 in M20, $\sim$ 1.7 in L20, and $\sim$ 2.7 in L30. The $\gamma$ increases with the collision speed for the same combination of colliding clouds. 
This tendency is similar to hydrodynamic simulation results of CCC performed by \citet{2018PASJ...70S..58T}.
It should be noted that for the same collision speed, 20 km $^{-1}$, $\gamma$ decreases with an increase in cloud size, as
$\gamma \sim$ 2.0 in M20 and $\gamma \sim$ 1.7 in L20. $\gamma  \sim$ 1.7 is observed for molecular clumps in molecular clouds \citep{Offner_2014}.

We show maps of H$_2$ column density, $N_{\mathrm{H}_2}$, viewed along collision axis with markers at the positions of massive bound cores and with unit vectors of the magnetic fields at $t$ = 3.1 Myr in all CCC models in Figure \ref{fig:column_CCC}.
We show the core mass in the unit of $M_{\odot}$ by a number on each marker. Here we exclude gas less than $\rho_0$ and assume particle number ratio of H$_{2}$ to He atoms of 4.7 (i.e., mean molecular weight $\mu$ = 2.35$m_{\mathrm{H}}$) in the calculation of column density. Here we assume that molecular clouds are dominated by H$_{2}$.
In all CCC models, there are major filaments roughly parallel to the $z$-axis near $y$ = 0 pc, and they are perpendicular to the initial magnetic fields. 
In M10 and L20, very massive bound cores greater than 100 $M_{\odot}$ are spatially associated with these filaments.
The magnetic field structures are shown by unit vectors which show directions of magnetic fields averaged by mass-weighted in the direction of the collision axis. The magnetic field directions are roughly perpendicular to the filaments in all CCC models.

We compare the results of CCC models with isolated cloud models (IM0 and IL0) shown in Appendix. CMDs in the isolated cloud models show that the formation of massive bound cores in those models is slow and less effective than M10 and L20.

\subsection{Comparison with \citet{2021PASJ...73S..75E} observational result}

We have demonstrated that there is an upper threshold of collision speed which allows formation of massive bound cores in CCCs, and it depends on colliding clouds, Small and Medium clouds and Small and Large clouds.
Our simulation results indicate that CCCs with higher collision speeds need a larger size of target cloud for massive bound core formation.
These results are similar to the observational results summarised by \citet{2021PASJ...73S..75E}.
They have reported a positive correlation between peak column densities of H$_{2}$ observed in colliding clouds with massive star formation and collision speeds of these clouds. 

We compare our simulation results with \citet{2021PASJ...73S..75E}.
We show time evolution of peak H$_2$ column densities in our numerical results in Figure \ref{fig:discuss}.
Top panel in Figure \ref{fig:discuss} shows time evolution of peak H$_2$ column density, $N_{\mathrm{H}_2}^{\mathrm{peak}}$, calculated along the collision axis with 1 pc spatial resolution in each CCC model.
After massive bound core formation, $N_{\mathrm{H}_2}^{\mathrm{peak}}$ increases gradually with time in M10 and L20.
The bottom panel in Figure \ref{fig:discuss} shows $N_{\mathrm{H}_2}^{\mathrm{peak}}$ at the epoch when Small cloud starts to move ($t$ = 0.5 Myr) and at the final simulation epoch ($t$ = 3.1 Myr) for the collision speeds in our CCC models on Figure 9(a) in \citet{2021PASJ...73S..75E}. For $t$ = 0.5 Myr, we show initial collision speeds. For $t$ = 3.1 Myr, we estimate collision speeds as half of (max.($v_x$) + avg.($v_x$)), where max($v_x$) and avg.($v_x$) are maximum values and density-weighted averages of x-component of velocity of gas with density more than $\rho_0$, respectively. The values of max($v_x$) and avg.($v_x$) are used as upper and lower ends of the error bars. Here, we mimic a  velocity estimation method used in observations.
We also show the total number of massive bound cores at $t$ = 3.1 Myr beside markers of our CCC models. Figure 9(a) in
\citet{2021PASJ...73S..75E} shows the peak column densities and collision velocities of observed CCCs with colored markers. Marker color shows the number of OB stars observed in the CCCs. They suggested a positive correlation between the peak column densities and collision velocities in observed CCCs. 
It implies that a collision with a higher relative velocity requires a higher column density to trigger star formation by CCC. Our simulation results show that $N_{\mathrm{H}_2}^{\mathrm{peak}}$ changes by a factor of 1.4 $\sim $ 3.4 during CCC in each CCC model, and larger cloud models have larger $N_{\mathrm{H}_2}^{\mathrm{peak}}$. For same collision speed, number of massive bound cores increases with $N_{\mathrm{H}_2}^{\mathrm{peak}}$. The higher speed models for the same colliding clouds reduce number of massive bound cores. This implies that the maximum collision speeds of CCCs for massive bound core formation increase with the peak column density. 
The maximum collision speeds are higher than the collision speeds obtained in the observed CCCs for the same maximum column density. Since we can expect massive star formation in CCCs with higher column density than our simulation model for the same collision speed as discussed in the following paragraphs, the upper limit of collision speed is consistent with \citet{2021PASJ...73S..75E}, if a massive star is formed in a massive bound core. It should be noted that there is some ambiguity in such a comparison of observations with simulations. The initial collision speeds of our models can be higher than the observed collision speeds of CCCs due to the effect of projection. Also, mean shock speeds at $t$ = 3.1 Myr in our models are not physically same as observed collision speeds, but they give us rough estimates of collision speeds at this epoch. $N_{\mathrm{H}_2}^{\mathrm{peak}}$ in our numerical results is consistent with the observed CCCs with small number of OB stars in \citet{2021PASJ...73S..75E}. CCC simulations that correspond to formation of a larger number of OB stars should be done to understand the correlation given in \citet{2021PASJ...73S..75E}. 
For such a numerical simulation, we need more detailed information of observed colliding clouds which show formation of massive star clusters, e.g., masses and sizes of colliding clouds, their density structures, their inner turbulence, and magnetic fields.

We estimate column density of a shocked layer with massive star formation in colliding clouds. This estimation is for comparison with our CCC models and with the observed correlation of column density, $N_{\textrm{obs}}$, and collision speed, $v_0$, which can be approximated as $N_{\textrm{obs}} \propto v_0^{3.4}$ as shown in \citet{2021PASJ...73S..75E}.
 
The gravitational unstable condition of the shocked layer induced by CCCs is proposed as,
\begin{equation}\label{eq:dis1}
\frac{l}{v_1} > \sqrt{\frac{3 \pi}{32 G \rho_{2}}},
\end{equation}
where $l$ is a typical scale of clouds swept by a shock induced by CCC, $v_1$ is collision velocity of CCC, and $\rho_{2}$ is density of the shocked layer. In this equation, we assume that duration of collision should be longer than the free-fall time of the shocked layer to be gravitational unstable. We assume a simple shock condition as
\begin{equation}\label{eq:dis2}
\rho_{1} v_1^{2}=\frac{B_{2}^{2}}{8 \pi},
\end{equation}
where $\rho_{1}$ is gas density of the clouds before the collision and $B_{2}$ is a magnetic field in the shocked layer and is assumed to dominate pressure in the shocked layer. $B_{2}$ is simply given as
\begin{equation}\label{eq:dis3}
B_{2} = B_{1}\left(\rho_{2} / \rho_{1}\right)
\end{equation}
for a one dimensional compression of magnetic field, $B_1$, perpendicular to the shocked layer, where $B_{1}$ is magnetic field in the pre-shock gas. From equations (\ref{eq:dis1}), (\ref{eq:dis2}), and (\ref{eq:dis3}), we estimate minimum column density required for gravitational instability of shocked layer as
\begin{equation}\label{eq:dis6}
N=\rho_{1} l>\sqrt{\frac{3 \pi B_{1} v_1 \sqrt{\rho_{1}}}{32 G \sqrt{8 \pi}}}.
\end{equation}
From this, we have
\begin{equation}\label{eq:dis7}
\begin{aligned}
N_\textrm{H$_2$}> 1.7\times10^{21}&\left(\frac{v_1}{10 \textrm{ km s$^{-1}$}}\right)^{0.5}\left(\frac{B_{1}}{4 \textrm{ \textmu G}}\right)^{0.5}\\
                  & \times\left(\frac{\rho_{1}}{3.67\times10^{-22} \textrm{ g cm$^{-3}$}}\right)^{0.25} \textrm{ cm$^{-2}$}.
\end{aligned}       
\end{equation}
This equation shows a column density of the gravitational unstable shocked layer is larger than the threshold value that is proportional to $v^{0.5}$, ${\rho_{1}}^{0.25}$, and ${B_{1}}^{0.5}$.
This is a necessary condition for massive star formation. This should be a condition for bound core formation. Massive stars may form in the shocked layer which satisfy this condition.
Further condition is needed for formation of massive bound core. 
The dotted line in Figure \ref{fig:discuss} shows this condition given by equation (\ref{eq:dis7}), using values of $\rho_{1}$ and $B_{1}$ same as the initial values in our clouds. Our results are consistent with this condition, since bound cores are formed in all CCC models. The observed CCCs by \citet{2021PASJ...73S..75E} satisfy the condition of equation (\ref{eq:dis7}). The observed colliding clouds in the range of high peak column density in Figure \ref{fig:discuss} are found in the galactic central region where observed clouds have higher magnetic field strengths and higher gas densities than our initial cloud models \citep{2019A&A...630A..74M,2017arXiv170505332M}. The dashed line in Figure \ref{fig:discuss} shows the condition given by equation (\ref{eq:dis7}) for higher values of $\rho_{1}$ and $B_{1}$ as 100$\rho_0$ and 1 mG which correspond to the observed values in molecular clouds in the galactic central region. This condition is well consistent with the observed colliding clouds in the range of high peak column density. 
Further study using colliding cloud models similar to the observed ones in the galactic central region is needed to understand the observed correlation given by \citet{2021PASJ...73S..75E}.

\subsection{Role of magnetic fields (Comparison with Paper I)}
In this paper, we have shown how collision speed controls massive bound core formation in colliding clouds with magnetic field and the magnetic field has the important role in this process, as summarized in section \ref{discuss_1_cols}. 
In the models in which massive bound cores are not formed before the shock crossed epoch of the target cloud, mass growth of massive cores is suppressed by the expansion of the shocked region which occurs after the shock crossed epoch of target cloud. This expansion is induced by the magnetic fields enhanced by the shock compression and such expansion is not found in the hydrodynamic simulations \citep{2014ApJ...792...63T,2018PASJ...70S..58T}.

In Paper I, we have shown the role of magnetic field which suppresses the non-linear thin shell in some smaller scale and promotes massive bound core formation. 
The reason for these opposite roles of magnetic field in Paper I and current paper is as follows.

In Paper I, we used colliding clouds of M10 model with 10 km s$^{-1}$ collision speed with various directions of magnetic fields and we found that massive bound cores form well in the duration of collision before the expansion of the shocked region.
In this paper, we have used higher collision speeds than Paper I.
If the duration of collision is not enough for massive bound core formation by the higher collision speed as in M20, L30, and L40 models, the expansion of the shocked region by enhanced magnetic pressure can suppress massive bound core formation. If the duration of collision is long enough for massive bound core formation by increasing of the target cloud size as in L20 model, the expansion of the shocked region occurs after massive bound core formation. The relation between the duration time of CCCs and the massive core formation time scale controls numerical results in this paper.

\subsection{Future works}
We plan on extending our simulations to CCCs with higher column density and with stronger magnetic fields.
\citet{2021PASJ...73S..75E} reported that higher peak column densities than our simulation models are observed in colliding clouds which are associated with super star clusters and active star formation regions in the central region of our Galaxy. 
It is interesting to study how many massive stars can be formed by CCCs with observed high peak column density.
Since strong magnetic fields of the order of 1 mG are estimated in the observed molecular clouds in the central molecular zone of our Galaxy \citep{2019A&A...630A..74M}, the effect of such a strong magnetic field on massive star formation process by CCCs should be investigated. 

In this paper, we do not consider feedback effects by newly formed massive stars by CCCs. 
Such stellar feedback should affect core formation during CCC. Hydrodynamic simulations of CCCs with the stellar feedback by photoionization were performed by \citet{2018PASJ...70S..54S}, and they show that the feedback can promote massive star formation during  CCCs. Stellar feedback should be included in future MHD CCC simulations, since the massive star formation is observed in several CCCs in the central regions of our Galaxy where magnetic fields are very strong. We will address this in our future papers.

\section{Summary}\label{summmary}
We study the effects of the magnetic fields on the formation of massive, self-gravitationally bound cores (MBCs) in high-speed cloud-cloud collisions (CCCs). Extending our previous work \citep{2021PASJ...73S.385S}, we perform magnetohydrodynamic simulations of these high-speed CCCs to make clear the role of collision speed on these processes.
We assumed two combinations of magnetized, colliding clouds, Small and Medium clouds (referred with M), and Small and Large clouds (referred with L). These clouds initially have the typical density of giant molecular clouds. Their cloud sizes are 7 - 20 pc, their masses are 10$^3$ - 10$^4$ $M_{\odot}$, and they have the internal turbulence.
The clouds are immersed in the uniform magnetic field of $B_0$ = 4.0 \textmu G normal to the collision axis. The magnetic fields are modified by the internal turbulent motion in the clouds. 
The collision speeds of 10 and 20 km s$^{-1}$ are assumed for Small and Medium clouds (M10 and M20) and 20, 30, and 40 km s$^{-1}$ are assumed for Small and Large clouds (L20, L30, and L40). We summarize our numerical results as follows.

\begin{enumerate}
\renewcommand{\labelenumi}{\arabic{enumi}.}
\item In models M20, L30, and L40, the MBC formation hardly occurs. In these models, the duration of the collision is relatively short in the sense that the cores cannot sufficiently accrete the gas to get gravitationally bound in the course of the collision. After the duration of the collision, the shocked region containing the cores rapidly expands to the ambient medium. Such expansion is driven by the magnetic pressure much higher than the ram pressure of the ambient medium. The expansion leads to the destruction of the unbound cores and suppression of core mass growth and MBC formation.

\item In models M10 and L20, on the other hand, the duration of the collision is relatively long, and the MBC formation efficiently occurs. The cores become tightly gravitationally bound while in the shocked layer created at the interface between the colliding clouds. The cores remain bound even while the shocked region expands into the ambient medium after the collision.

\item Overall, the MBC formation is inefficient in the former models, M20, L30, and L40, while efficient in the latter models, M10 and L20. These results indicate the critical speeds above which the magnetic fields contribute to suppressing the MBC formation, depending on different combinations of the colliding clouds. No previous hydrodynamic simulations of CCC models have reported this negative effect of magnetic fields on the MBC formation.

\item Together with our previous work, we conclude that the magnetic fields provide the two competing effects on the formation of MBCs in CCC. They promote the mass accumulation into cores during a collision, whereas they operate to destroy cores or hinder the core growth after the collision. The duration of collision determines which effect prevails, providing the maximum collision speed for the MBC formation for given colliding clouds.

\item Our results indicate that the MBC formation in colliding clouds with a higher speed requires a higher initial column density of the colliding clouds and that the maximum collision speed depends on column density of colliding clouds. These properties are very similar to observed properties of colliding clouds with OB stars reported by \citet{2021PASJ...73S..75E}.
\end{enumerate}
\section*{Acknowledgements}

The authors thank Tsuyoshi Inoue, Kazuo Sorai, and Shu-ichiro Inutsuka for their fruitful discussions. Numerical computations were carried out on the Cray XC50 supercomputer systems at the Center for Computational Astrophysics of the National Astronomical Observatory of Japan. Numerical analysis was done using the yt, a data analysis and visualization package \citep{2011ApJS..192....9T}. NS was supported by JST SPRING Grant (JPMJSP2119). AH is funded by the JSPS KAKENHI Grant (JP19K03923). TO is supported by JSPS KAKENHI Grants (19H01931, 20H05861, 21H04496). TH appreciates the support by JSPS KAKENHI Grants (17H06360, 19H01934, 21H00041).




\bibliographystyle{mnras}



\bibliography{reference.bib}


\appendix
\section{Isolated clouds} \label{isolated} 
We show numerical results of the isolated cloud models (IM0 and IL0) and compare them with those of CCC models shown in main text.

\begin{figure}
    \begin{center}
    \includegraphics[width=0.5\textwidth]{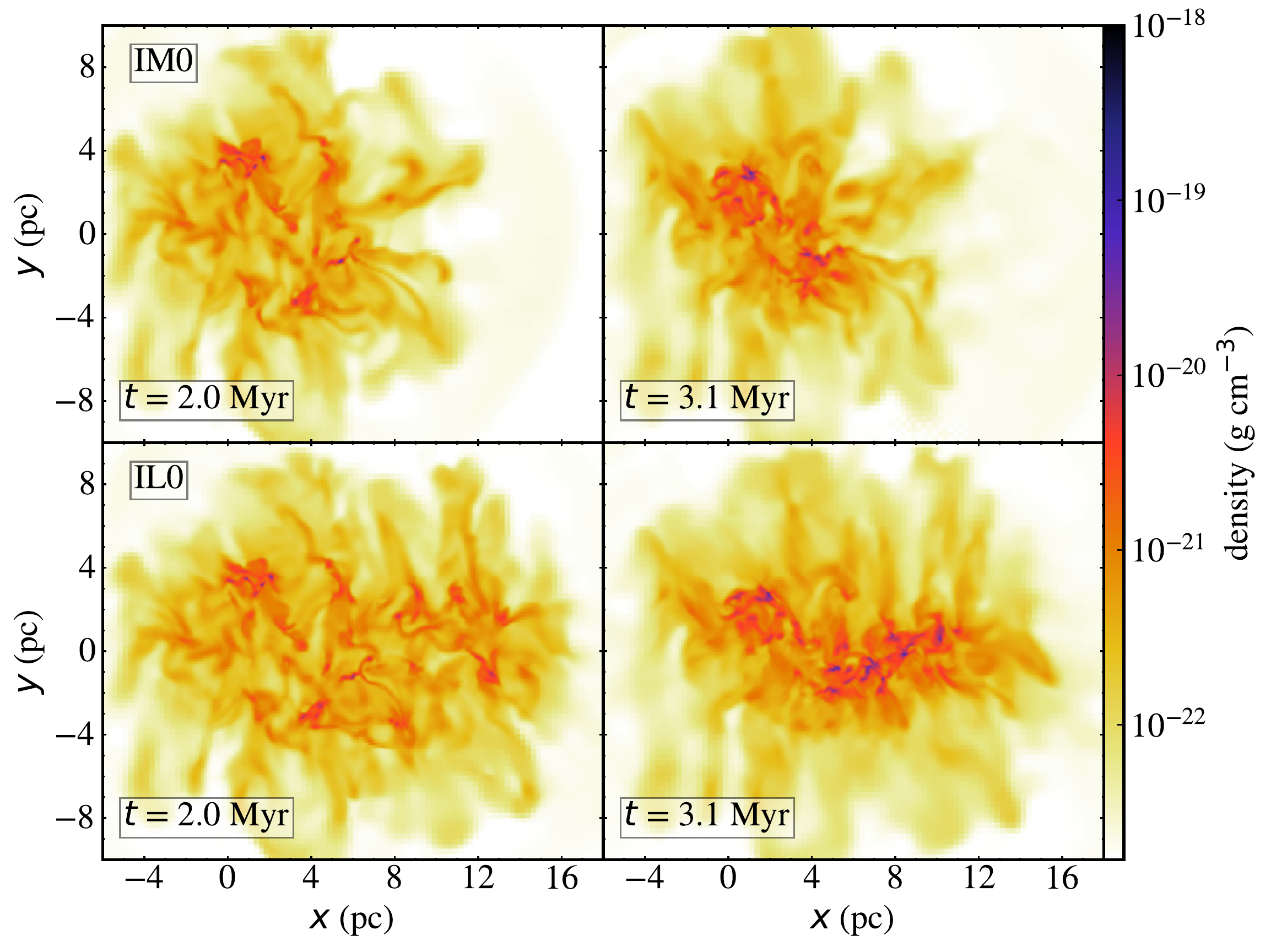}
    \end{center}
    \caption{Slice plots of the gas density in $z$ = 0 pc at $t$ = 2.0 (left panels) and 3.1 Myr (right panels) in the IM0 (top panels) and IL0 (bottom panels). The color bar shows gas density.} 
    \label{fig:slices_isolated}
\end{figure}

The density slice plots at $t$ = 2.0 and 3.1 Myr for IM0 (top panels) and IL0 (bottom panels) are shown in Figure \ref{fig:slices_isolated}. This figure shows a clear difference between the dense regions in the isolated cloud models and CCC models. More dense regions are formed in CCC models than the isolated cloud models.

We show time evolution of $M_{\mathrm{high,tot}}$ in IM0 and IL0 in right panel of Figure \ref{fig:highdenseCCC}.
This figure shows that $M_{\mathrm{high,tot}}$ in IL0 and IM0 increases much slower than the CCC models. By $t$ = 3.1 Myr, growth of $M_{\mathrm{high,tot}}$ is lower in the isolated cloud models than M10 and L20.

\begin{figure}
    \begin{center}
    \includegraphics[width=0.5\textwidth]{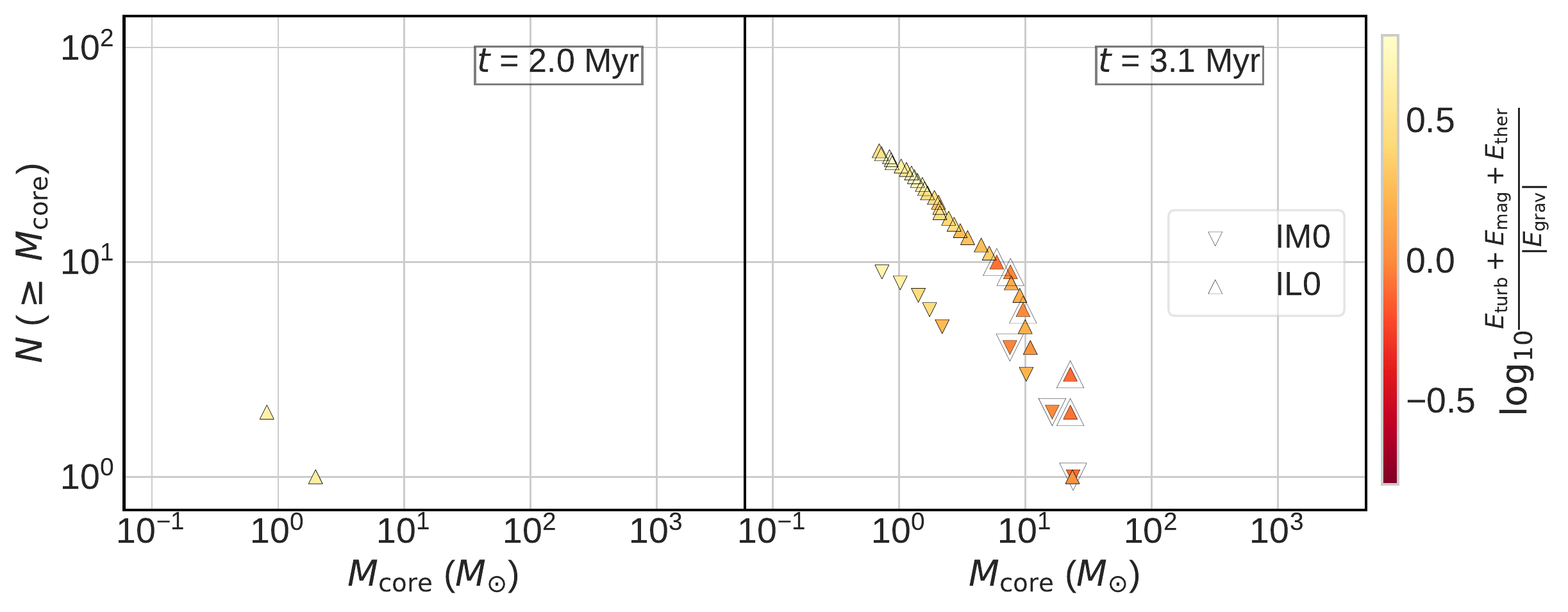}
    \end{center}
    \caption{Cumulative core mass distributions shown by filled down-pointing and up-pointing triangles at $t$ = 2.0 (top panel) and 3.1 Myr (bottom panel) for the IM0 and IL0, respectively. The gravitationally bound cores are marked by larger, open down-pointing and up-pointing triangles for the IM0 and IL0, respectively.}
    \label{fig:cmf_isolated}
\end{figure}

We show CMDs of both models at $t$ = 2.0 and 3.1 Myr in Figure \ref{fig:cmf_isolated}. This figure shows that dense core formation occurs later than CCC models.
At $t$ = 3.1 Myr, there are two massive bound cores in both models and a significantly greater number of dense cores in IL0 than IM0. 
These numbers of massive bound cores and maximum masses of dense cores in isolated cloud models are much smaller than those in M10 and L20. 

\bsp	
\label{lastpage}
\end{document}